\documentclass[twocolumn,tighten]{aastex62}

\usepackage{newtxtext,newtxmath}

\usepackage[T1]{fontenc}
\usepackage{ae,aecompl}

\usepackage{graphicx}	
\usepackage{amsmath}	
\usepackage{amssymb}	
\usepackage{enumerate}
\usepackage{xcolor}

\DeclareMathAlphabet{\mathcal}{OMS}{cmsy}{m}{n}

\definecolor{forestgreen}{rgb}{0.13, 0.75, 0.13}

\newcommand{\kmsmpc}{{\rm km}\,{\rm s}^{-1}\,{\rm Mpc}^{-1}}
\newcommand{\cov}{\mathcal{C}_\eta}
\newcommand{\Z}{\mathcal{Z}}

\defcitealias{Guy2010}{G10}
\defcitealias{Chotard2011}{C11}
\defcitealias{Rubin2015}{R15}	
\defcitealias{Scolnic2016}{SK16}	 
\newcommand{\gten}{\citetalias{Guy2010}}
\newcommand{\steve}{\textit{Steve}}
\newcommand{\celeven}{\citetalias{Chotard2011}}
\newcommand{\rubin}{\citetalias{Rubin2015}}
\newcommand{\sk}{\citetalias{Scolnic2016}}

\shorttitle{{\steve}}
\begin{document}
\title{{\steve}: A hierarchical Bayesian model for Supernova Cosmology}

\author{S.~R.~Hinton}
\affiliation{School of Mathematics and Physics, University of Queensland,  Brisbane, QLD 4072, Australia}
\author{T.~M.~Davis}
\affiliation{School of Mathematics and Physics, University of Queensland,  Brisbane, QLD 4072, Australia}
\author{A.~G.~Kim}
\affiliation{Lawrence Berkeley National Laboratory, 1 Cyclotron Road, Berkeley, CA 94720, USA}
\author{D.~Brout}
\affiliation{Department of Physics and Astronomy, University of Pennsylvania, Philadelphia, PA 19104, USA}
\author{C.~B.~D'Andrea}
\affiliation{Department of Physics and Astronomy, University of Pennsylvania, Philadelphia, PA 19104, USA}
\author{R.~Kessler}
\affiliation{Department of Astronomy and Astrophysics, University of Chicago, Chicago, IL 60637, USA}
\affiliation{Kavli Institute for Cosmological Physics, University of Chicago, Chicago, IL 60637, USA}
\author{J.~Lasker}
\affiliation{Department of Astronomy and Astrophysics, University of Chicago, Chicago, IL 60637, USA}
\affiliation{Kavli Institute for Cosmological Physics, University of Chicago, Chicago, IL 60637, USA}
\author{C.~Lidman}
\affiliation{The Research School of Astronomy and Astrophysics, Australian National University, ACT 2601, Australia}
\author{E.~Macaulay}
\affiliation{Institute of Cosmology and Gravitation, University of Portsmouth, Portsmouth, PO1 3FX, UK}
\author{A.~M\"oller}
\affiliation{ARC Centre of Excellence for All-sky Astrophysics (CAASTRO)}
\affiliation{The Research School of Astronomy and Astrophysics, Australian National University, ACT 2601, Australia}
\author{M.~Sako}
\affiliation{Department of Physics and Astronomy, University of Pennsylvania, Philadelphia, PA 19104, USA}
\author{D.~Scolnic}
\affiliation{Kavli Institute for Cosmological Physics, University of Chicago, Chicago, IL 60637, USA}
\author{M.~Smith}
\affiliation{School of Physics and Astronomy, University of Southampton,  Southampton, SO17 1BJ, UK}
\author{R.~C.~Wolf}
\affiliation{Graduate School of Education, Stanford University, 160, 450 Serra Mall, Stanford, CA 94305, USA}
\author{M.~Childress}
\affiliation{School of Physics and Astronomy, University of Southampton,  Southampton, SO17 1BJ, UK}
\author{E.~Morganson}
\affiliation{National Center for Supercomputing Applications, 1205 West Clark St., Urbana, IL 61801, USA}
\author{S.~Allam}
\affiliation{Fermi National Accelerator Laboratory, P. O. Box 500, Batavia, IL 60510, USA}
\author{J.~Annis}
\affiliation{Fermi National Accelerator Laboratory, P. O. Box 500, Batavia, IL 60510, USA}
\author{S.~Avila}
\affiliation{Institute of Cosmology and Gravitation, University of Portsmouth, Portsmouth, PO1 3FX, UK}
\author{E.~Bertin}
\affiliation{CNRS, UMR 7095, Institut d'Astrophysique de Paris, F-75014, Paris, France}
\affiliation{Sorbonne Universit\'es, UPMC Univ Paris 06, UMR 7095, Institut d'Astrophysique de Paris, F-75014, Paris, France}
\author{D.~Brooks}
\affiliation{Department of Physics \& Astronomy, University College London, Gower Street, London, WC1E 6BT, UK}
\author{D.~L.~Burke}
\affiliation{Kavli Institute for Particle Astrophysics \& Cosmology, P. O. Box 2450, Stanford University, Stanford, CA 94305, USA}
\affiliation{SLAC National Accelerator Laboratory, Menlo Park, CA 94025, USA}
\author{A.~Carnero~Rosell}
\affiliation{Centro de Investigaciones Energ\'eticas, Medioambientales y Tecnol\'ogicas (CIEMAT), Madrid, Spain}
\affiliation{Laborat\'orio Interinstitucional de e-Astronomia - LIneA, Rua Gal. Jos\'e Cristino 77, Rio de Janeiro, RJ - 20921-400, Brazil}
\author{M.~Carrasco~Kind}
\affiliation{Department of Astronomy, University of Illinois at Urbana-Champaign, 1002 W. Green Street, Urbana, IL 61801, USA}
\affiliation{National Center for Supercomputing Applications, 1205 West Clark St., Urbana, IL 61801, USA}
\author{J.~Carretero}
\affiliation{Institut de F\'{\i}sica d'Altes Energies (IFAE), The Barcelona Institute of Science and Technology, Campus UAB, 08193 Bellaterra (Barcelona) Spain}
\author{C.~E.~Cunha}
\affiliation{Kavli Institute for Particle Astrophysics \& Cosmology, P. O. Box 2450, Stanford University, Stanford, CA 94305, USA}
\author{L.~N.~da Costa}
\affiliation{Laborat\'orio Interinstitucional de e-Astronomia - LIneA, Rua Gal. Jos\'e Cristino 77, Rio de Janeiro, RJ - 20921-400, Brazil}
\affiliation{Observat\'orio Nacional, Rua Gal. Jos\'e Cristino 77, Rio de Janeiro, RJ - 20921-400, Brazil}
\author{C.~Davis}
\affiliation{Kavli Institute for Particle Astrophysics \& Cosmology, P. O. Box 2450, Stanford University, Stanford, CA 94305, USA}
\author{J.~De~Vicente}
\affiliation{Centro de Investigaciones Energ\'eticas, Medioambientales y Tecnol\'ogicas (CIEMAT), Madrid, Spain}
\author{D.~L.~DePoy}
\affiliation{George P. and Cynthia Woods Mitchell Institute for Fundamental Physics and Astronomy, and Department of Physics and Astronomy, Texas A\&M University, College Station, TX 77843,  USA}
\author{P.~Doel}
\affiliation{Department of Physics \& Astronomy, University College London, Gower Street, London, WC1E 6BT, UK}
\author{T.~F.~Eifler}
\affiliation{Department of Astronomy/Steward Observatory, 933 North Cherry Avenue, Tucson, AZ 85721-0065, USA}
\affiliation{Jet Propulsion Laboratory, California Institute of Technology, 4800 Oak Grove Dr., Pasadena, CA 91109, USA}
\author{B.~Flaugher}
\affiliation{Fermi National Accelerator Laboratory, P. O. Box 500, Batavia, IL 60510, USA}
\author{P.~Fosalba}
\affiliation{Institut d'Estudis Espacials de Catalunya (IEEC), 08034 Barcelona, Spain}
\affiliation{Institute of Space Sciences (ICE, CSIC),  Campus UAB, Carrer de Can Magrans, s/n,  08193 Barcelona, Spain}
\author{J.~Frieman}
\affiliation{Fermi National Accelerator Laboratory, P. O. Box 500, Batavia, IL 60510, USA}
\affiliation{Kavli Institute for Cosmological Physics, University of Chicago, Chicago, IL 60637, USA}
\author{J.~Garc\'ia-Bellido}
\affiliation{Instituto de Fisica Teorica UAM/CSIC, Universidad Autonoma de Madrid, 28049 Madrid, Spain}
\author{E.~Gaztanaga}
\affiliation{Institut d'Estudis Espacials de Catalunya (IEEC), 08034 Barcelona, Spain}
\affiliation{Institute of Space Sciences (ICE, CSIC),  Campus UAB, Carrer de Can Magrans, s/n,  08193 Barcelona, Spain}
\author{D.~W.~Gerdes}
\affiliation{Department of Astronomy, University of Michigan, Ann Arbor, MI 48109, USA}
\affiliation{Department of Physics, University of Michigan, Ann Arbor, MI 48109, USA}
\author{R.~A.~Gruendl}
\affiliation{Department of Astronomy, University of Illinois at Urbana-Champaign, 1002 W. Green Street, Urbana, IL 61801, USA}
\affiliation{National Center for Supercomputing Applications, 1205 West Clark St., Urbana, IL 61801, USA}
\author{J.~Gschwend}
\affiliation{Laborat\'orio Interinstitucional de e-Astronomia - LIneA, Rua Gal. Jos\'e Cristino 77, Rio de Janeiro, RJ - 20921-400, Brazil}
\affiliation{Observat\'orio Nacional, Rua Gal. Jos\'e Cristino 77, Rio de Janeiro, RJ - 20921-400, Brazil}
\author{G.~Gutierrez}
\affiliation{Fermi National Accelerator Laboratory, P. O. Box 500, Batavia, IL 60510, USA}
\author{W.~G.~Hartley}
\affiliation{Department of Physics \& Astronomy, University College London, Gower Street, London, WC1E 6BT, UK}
\affiliation{Department of Physics, ETH Zurich, Wolfgang-Pauli-Strasse 16, CH-8093 Zurich, Switzerland}
\author{D.~L.~Hollowood}
\affiliation{Santa Cruz Institute for Particle Physics, Santa Cruz, CA 95064, USA}
\author{K.~Honscheid}
\affiliation{Center for Cosmology and Astro-Particle Physics, The Ohio State University, Columbus, OH 43210, USA}
\affiliation{Department of Physics, The Ohio State University, Columbus, OH 43210, USA}
\author{E.~Krause}
\affiliation{Department of Astronomy/Steward Observatory, 933 North Cherry Avenue, Tucson, AZ 85721-0065, USA}
\author{K.~Kuehn}
\affiliation{Australian Astronomical Optics, Macquarie University, North Ryde, NSW 2113, Australia}
\author{N.~Kuropatkin}
\affiliation{Fermi National Accelerator Laboratory, P. O. Box 500, Batavia, IL 60510, USA}
\author{O.~Lahav}
\affiliation{Department of Physics \& Astronomy, University College London, Gower Street, London, WC1E 6BT, UK}
\author{M.~Lima}
\affiliation{Departamento de F\'isica Matem\'atica, Instituto de F\'isica, Universidade de S\~ao Paulo, CP 66318, S\~ao Paulo, SP, 05314-970, Brazil}
\affiliation{Laborat\'orio Interinstitucional de e-Astronomia - LIneA, Rua Gal. Jos\'e Cristino 77, Rio de Janeiro, RJ - 20921-400, Brazil}
\author{M.~A.~G.~Maia}
\affiliation{Laborat\'orio Interinstitucional de e-Astronomia - LIneA, Rua Gal. Jos\'e Cristino 77, Rio de Janeiro, RJ - 20921-400, Brazil}
\affiliation{Observat\'orio Nacional, Rua Gal. Jos\'e Cristino 77, Rio de Janeiro, RJ - 20921-400, Brazil}
\author{M.~March}
\affiliation{Department of Physics and Astronomy, University of Pennsylvania, Philadelphia, PA 19104, USA}
\author{J.~L.~Marshall}
\affiliation{George P. and Cynthia Woods Mitchell Institute for Fundamental Physics and Astronomy, and Department of Physics and Astronomy, Texas A\&M University, College Station, TX 77843,  USA}
\author{F.~Menanteau}
\affiliation{Department of Astronomy, University of Illinois at Urbana-Champaign, 1002 W. Green Street, Urbana, IL 61801, USA}
\affiliation{National Center for Supercomputing Applications, 1205 West Clark St., Urbana, IL 61801, USA}
\author{R.~Miquel}
\affiliation{Instituci\'o Catalana de Recerca i Estudis Avan\c{c}ats, E-08010 Barcelona, Spain}
\affiliation{Institut de F\'{\i}sica d'Altes Energies (IFAE), The Barcelona Institute of Science and Technology, Campus UAB, 08193 Bellaterra (Barcelona) Spain}
\author{R.~L.~C.~Ogando}
\affiliation{Laborat\'orio Interinstitucional de e-Astronomia - LIneA, Rua Gal. Jos\'e Cristino 77, Rio de Janeiro, RJ - 20921-400, Brazil}
\affiliation{Observat\'orio Nacional, Rua Gal. Jos\'e Cristino 77, Rio de Janeiro, RJ - 20921-400, Brazil}
\author{A.~A.~Plazas}
\affiliation{Jet Propulsion Laboratory, California Institute of Technology, 4800 Oak Grove Dr., Pasadena, CA 91109, USA}
\author{E.~Sanchez}
\affiliation{Centro de Investigaciones Energ\'eticas, Medioambientales y Tecnol\'ogicas (CIEMAT), Madrid, Spain}
\author{V.~Scarpine}
\affiliation{Fermi National Accelerator Laboratory, P. O. Box 500, Batavia, IL 60510, USA}
\author{R.~Schindler}
\affiliation{SLAC National Accelerator Laboratory, Menlo Park, CA 94025, USA}
\author{M.~Schubnell}
\affiliation{Department of Physics, University of Michigan, Ann Arbor, MI 48109, USA}
\author{S.~Serrano}
\affiliation{Institut d'Estudis Espacials de Catalunya (IEEC), 08034 Barcelona, Spain}
\affiliation{Institute of Space Sciences (ICE, CSIC),  Campus UAB, Carrer de Can Magrans, s/n,  08193 Barcelona, Spain}
\author{I.~Sevilla-Noarbe}
\affiliation{Centro de Investigaciones Energ\'eticas, Medioambientales y Tecnol\'ogicas (CIEMAT), Madrid, Spain}
\author{M.~Soares-Santos}
\affiliation{Brandeis University, Physics Department, 415 South Street, Waltham MA 02453}
\author{F.~Sobreira}
\affiliation{Instituto de F\'isica Gleb Wataghin, Universidade Estadual de Campinas, 13083-859, Campinas, SP, Brazil}
\affiliation{Laborat\'orio Interinstitucional de e-Astronomia - LIneA, Rua Gal. Jos\'e Cristino 77, Rio de Janeiro, RJ - 20921-400, Brazil}
\author{E.~Suchyta}
\affiliation{Computer Science and Mathematics Division, Oak Ridge National Laboratory, Oak Ridge, TN 37831}
\author{G.~Tarle}
\affiliation{Department of Physics, University of Michigan, Ann Arbor, MI 48109, USA}
\author{D.~Thomas}
\affiliation{Institute of Cosmology and Gravitation, University of Portsmouth, Portsmouth, PO1 3FX, UK}
\author{V.~Vikram}
\affiliation{Argonne National Laboratory, 9700 South Cass Avenue, Lemont, IL 60439, USA}
\author{Y.~Zhang}
\affiliation{Fermi National Accelerator Laboratory, P. O. Box 500, Batavia, IL 60510, USA}

\begin{abstract}
We present a new Bayesian hierarchical model (BHM) named {\steve} for performing type Ia supernova (SN~Ia) cosmology fits. 
This advances previous works by including 
an improved treatment of Malmquist bias, 
accounting for additional sources of systematic uncertainty, 
and increasing numerical efficiency. 
Given light curve fit parameters, redshifts, and host-galaxy masses, we fit {\steve} simultaneously for parameters describing cosmology, SN~Ia populations, and systematic uncertainties. Selection effects are characterised using Monte-Carlo simulations.
We demonstrate its implementation by fitting realisations of SN~Ia datasets where the SN~Ia model closely follows that used in {\steve}.
Next, we validate on more realistic SNANA simulations of SN~Ia samples from the Dark Energy Survey and low-redshift surveys \citep{DESKEY}.
These simulated datasets contain more than $60\,000$ SNe~Ia, which we use to evaluate biases in the recovery of cosmological parameters, specifically the equation-of-state of dark energy, $w$. 
This is the most rigorous test of a BHM method applied to SN~Ia cosmology fitting, and reveals small $w$-biases that depend on the simulated SN~Ia properties, in particular the intrinsic SN~Ia scatter model. This $w$-bias is less than $0.03$ on average, less than half the statistical uncertainty on $w$.
These simulation test results are a concern for BHM cosmology fitting applications on large upcoming surveys, and therefore future development will focus on minimising the sensitivity of {\steve} to the SN~Ia intrinsic scatter model.
\end{abstract}



\section{Introduction}

Two decades have passed since the discovery of the accelerating universe \citep{Riess1998, Perlmutter1999}. Since that time, the number of observed type Ia supernovae (SN~Ia) has increased by more than an order of magnitude, with contributions from modern surveys at both low redshift \citep{Bailey2008, Freedman2009, Hicken2009,  Contreras2010, Conley2011}, and higher redshift \citep{Astier2006, Wood-Vasey2007, Frieman2008, Balland2009, Amanullah2010, chambers2016panstarrs, sako2018sdss}. Cosmological analyses of these supernova samples \citep{Kowalski2008, Kessler2009, Conley2011, Suzuki2012, Betoule2014, Rest2014, Scolnic2017} have been combined with complementary probes of large scale structure and the CMB. For a recent review, see \citet{Huterer2018}. While these efforts have reduced the uncertainty on the equation-of-state of dark energy ($w$) by more than a factor of two, it is still consistent with a cosmological constant and the nature of dark energy remains an unsolved mystery.

In attempts to tease out the nature of dark energy, active and planned surveys are continually growing in size and scale. The Dark Energy Survey \citep[DES,][]{Bernstein2012, Abbott2016} has discovered thousands of type Ia supernovae, attaining both spectroscopically and photometrically identified samples. The Large Synoptic Survey Telescope \citep[LSST,][]{Ivezic2008, LSSTScienceCollaboration2009} will discover tens of thousands of photometrically classified supernovae. Such increased statistical power demands greater fidelity and flexibility in modelling supernovae for cosmological purposes, as we will require reduced systematic uncertainties to fully utilise these increased statistics \citep{Betoule2014, Scolnic2017}.

As such, considerable resources are aimed at developing more sophisticated supernova cosmology analyses. The role of simulations mimicking survey observations has become increasingly important in determining biases in cosmological constraints and validating specific supernova models. First used in SNLS \citep{Astier2006} and ESSENCE analyses \citep{Wood-Vasey2007}, and then refined and improved for SDSS \citep{Kessler2009}, simulations are a fundamental component of modern supernova cosmology.  \citet{Betoule2014} quantise and correct observational bias using simulations, and more recently \citet{Scolnic2016} and \citet{Kessler2017} explore simulations to quantify observational bias in SN~Ia distances as a function of multiple factors to improve bias correction. Approximate Bayesian computation methods also make use of simulations, trading traditional likelihoods and analytic approximations for more robust models with the cost of increased computational time \citep{Weyant2013, Jennings2016}. Bayesian Hierarchical models abound \citep{Mandel2009, March2011, March2014, Rubin2015, Shariff2016, Roberts2017}, and either use simulation-determined distance-corrections to correct for biases, or attempt to find analytic approximations for effects such as Malmquist bias to model the biases inside the BHM itself.

In this paper, we lay out a new hierarchical model that builds off the past work of \citet{Rubin2015}. We include additional sources of systematic uncertainty, including an analytic formulation of selection efficiency which incorporates parametric uncertainty. We also implement a different model of intrinsic dispersion to both incorporate redshift-dependent scatter and to increase numerical efficiency, allowing our model to perform rapid fits to supernovae datasets.

Section \ref{sec:review} is dedicated to a quick review of supernova cosmology analysis methods, and Section \ref{sec:challenges} outlines some of the common challenges faced by analysis methods. In Section \ref{sec:method} we outline our methodology. Model verification on simulated datasets is given in Section \ref{sec:verification}, along with details on potential areas of improvement. We summarise our methodology in Section \ref{sec:conclusion}.

\section{Review}
\label{sec:review}

Whilst supernova observations take the form of photometric time-series brightness measurements in many bands and a redshift measurement of the supernova (or its assumed host), most analyses do not work from these measurements directly. Instead, most techniques fit an observed redshift and these photometric observations to a supernova model, with the most widely used being that of the empirical SALT2 model \citep{Guy2007, Guy2010}. This model is trained separately before fitting the supernova light curves for cosmology \citep{Guy2010, Betoule2014}. The resulting output from the model is, for each supernova, an amplitude $x_0$ (which can be converted into apparent magnitude, $m_B = -2.5\log(x_0)$), a stretch term $x_1$, and color term $c$, along with a covariance matrix describing the uncertainty on these summary statistics ($\mathcal{C_\eta}$). As all supernovae are not identical, an ensemble of supernovae form a redshift-dependent, observed population of $\hat{m}_B$, $\hat{x}_1$ and $\hat{c}$, where the hat denotes an observed variable.

This ensemble of $\hat{m}_B$, $\hat{x}_1$ and $\hat{c}$ represents an observed population, which -- due to the presence of various selection effects -- may not represent the true, underlying supernova population. Accurately characterising this underlying population, its evolution over redshift, and effects from host-galaxy environment, is one of the challenges of supernova cosmology. Given some modelled underlying supernova population that lives in the redshift-dependent space $M_B$ (absolute magnitude of the supernova, traditionally in the Bessell $B$ band), $x_1$, and $c$, the introduction of cosmology into the model is simple -- it translates the underlying population from absolute magnitude space into the observed population in apparent magnitude space. Specifically, for any given supernova our map between absolute magnitude and apparent magnitude is given by the distance modulus:
\begin{equation}
\mu_{\rm obs} = m_B + \alpha x_1 - \beta c - M_B + \Delta M \cdot m + \text{other corrections}, \label{eq:standard}
\end{equation}
where $M_B$ is the mean absolute magnitude for all SN~Ia given $x_1=c=0$,  $\alpha$ is the stretch correction \citep{Phillips1993, Phillips1999}, and $\beta$ is the color correction \citep{Tripp1998} that respectively encapsulate the empirical relation that broader (longer-lasting) and bluer supernovae are brighter. $\Delta M \cdot m$ refers to the host-galaxy mass correlation discussed in Section \ref{sec:hostgal}. The distance modulus $\mu_{\rm obs}$ is a product of our observations, however a distance modulus $\mu_C$ can be precisely calculated given cosmological parameters and a redshift. The `$\text{other corrections}$' term often includes bias corrections for traditional $\chi^2$ analyses. Bias corrections can take multiple forms, such as a redshift-dependent function \citep{Betoule2014} or a 5D function of $c$, $x_1$, $\alpha$, $\beta$ and $z$ \citep{Kessler2017, Scolnic2017}.

\subsection{Traditional Cosmology Analyses} \label{sec:traditional}

Traditional $\chi^2$ analyses such as that found in \citet{Riess1998, Perlmutter1999, Wood-Vasey2007, Kowalski2008, Kessler2009, Conley2011, Betoule2014}, minimise the difference in distance modulus between the observed distance modulus attained from equation \ref{eq:standard}, and the cosmologically predicted distance modulus. The $\chi^2$ function minimised is
\begin{equation}
\chi^2 = (\mu_{\rm obs} - \mu_C)^\dagger C^{-1}_{\mu} (\mu_{\rm obs} - \mu_C),
\end{equation}
where $C^{-1}_{\mu}$ is an uncertainty matrix that combines statistical and systematic uncertainties (see \citet{Brout18SYS} for a review of these uncertainties for the DES supernova analysis). The predicted $\mu_C$ is defined as
\begin{eqnarray}
\mu_C &=& 5 \log\left[ \frac{d_L}{10{\rm pc}} \right], \label{eq:mu0}\\
d_L &=& (1+z)\frac{c}{H_0} \int_0^z \frac{dz'}{E(z')},\\
E(z) &=& \sqrt{ \Omega_m (1+z')^3 + \Omega_k (1 + z')^2 + \Omega_\Lambda (1+z')^{3(1+w)}}
\end{eqnarray}
where $d_L$ is the luminosity distance for redshift $z$ given a specific cosmology, $H_0$ is the current value of Hubble's constant in $\kmsmpc$ and $\Omega_m$, $\Omega_k$, and $\Omega_\Lambda$ represent the energy density terms for matter, curvature and dark energy respectively.

The benefit this analysis methodology provides is speed -- for samples of hundreds of supernovae or less, efficient matrix inversion algorithms allow the likelihood to be evaluated quickly. The speed comes with several limitations. Firstly, formulating a $\chi^2$ likelihood results in a loss of model flexibility by assuming Gaussian uncertainty. Secondly, the method of creating a covariance matrix relies on computing partial derivatives and thus any uncertainty estimated from this method loses information about correlation between sources of uncertainty. For example, the underlying supernova color population's mean and skewness are highly correlated, however this correlation is ignored when determining population uncertainty using numerical derivatives of population permutations. Whilst correlations can be incorporated into a covariance matrix, it requires human awareness of the correlations and thus methods that can automatically capture correlated uncertainties are preferable. Thirdly, the computational efficiency is dependent on both creating the global covariance matrix, and then inverting a covariance matrix with dimensionality linearly proportional to the number of supernovae. As this number increases, the cost of inversion rises quickly, and is not viable for samples with thousands of supernovae. A recent solution to this computational cost problem is to bin the supernovae in redshift bins, which produces a matrix of manageable size and can allow fitting without matrix inversion at every step. Whilst binning data results in some loss of information, recent works tested against simulations show that this loss does not result in significant cosmological biases \citep{Scolnic2016, Kessler2017}.

\begin{figure}
	\begin{center}
		\includegraphics[width=\columnwidth]{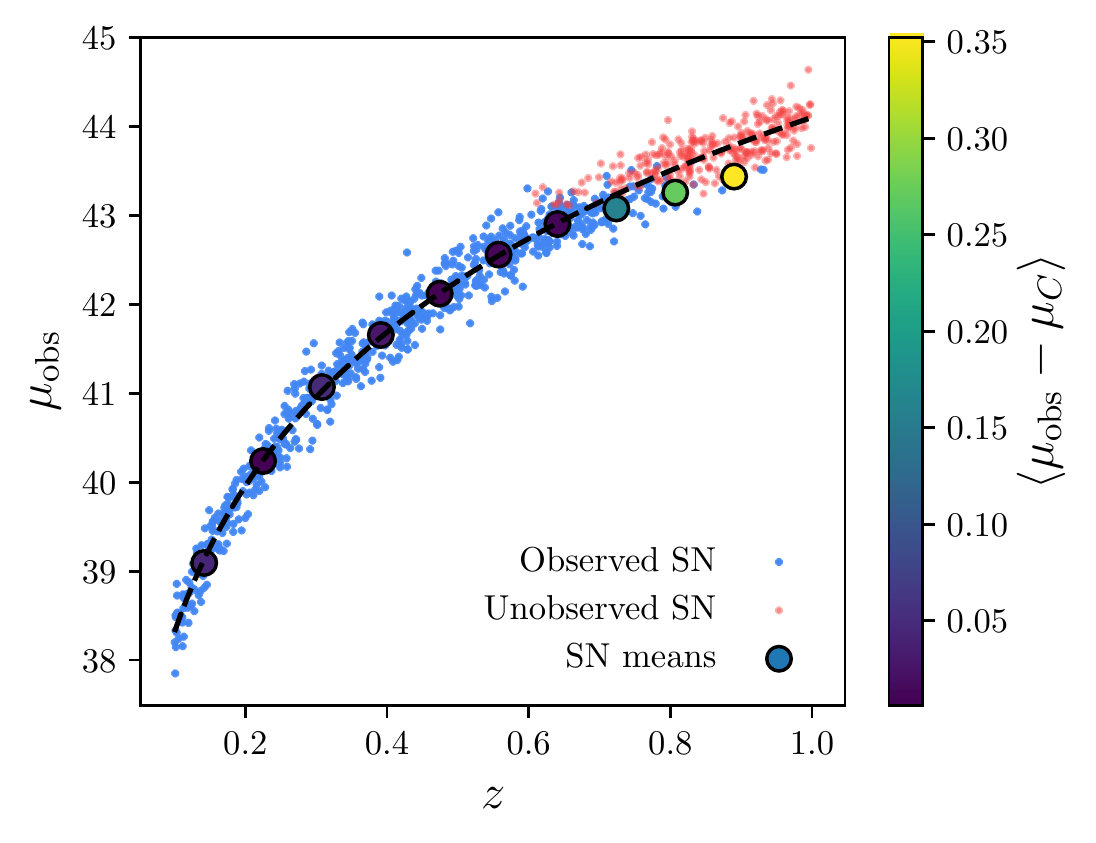}
	\end{center}
	\caption{An example of the effects of Malmquist bias. Here are shown 1000 simulated supernovae redshifts and distance modulus given fiducial cosmology. The simulated survey is magnitude limited, and all supernovae brighter than magnitude 24 are successfully observed (shown as blue dots), and all dimmer than 24th magnitude are not successfully observed (shown as red dots). By binning the supernovae along redshift, and taking the mean distance modulus of the supernovae in each bin, we can see that at higher redshift where Malmquist bias kicks in, the population mean drops and becomes biased. This source of bias must either be corrected by adjusting the data (such as subtracting the found bias) for by incorporating Malmquist bias explicitly in the cosmological model.}
	\label{fig:malmquist}
\end{figure}

Selection efficiency, such as the well known Malmquist bias \citep{MalmquistK.G.1922} is accounted for by correcting the determined $\mu_{\rm obs}$ from the data, or equivalently, adding a distance bias to the $\mu_C$ prediction. Specifically, Malmquist bias is the result of losing the fainter tail of the supernova population at high redshift. An example of Malmquist bias is illustrated in Figure \ref{fig:malmquist}, which simulates supernovae according to equation \eqref{eq:standard}. Simulations following survey observational strategies and geometry are used to calculate the expected bias in distance modulus, which is then subtracted from the observational data. When using traditional fitting methods such as that found in \citet{Betoule2014}, these effects are not built into the likelihood and instead are formed by correcting data. This means that the bias uncertainty is not captured fully in the $\chi^2$ distribution, and subtle correlations between cosmological or population parameters and the bias correction is lost. Recent developments such as the BBC method \citep{Kessler2017} incorporate corrections dependent on $\alpha$ and $\beta$, improving their capture of uncertainty on bias corrections in the $\chi^2$ likelihood.

\subsection{Approximate Bayesian Computation}

To avoid the limitations of the traditional approaches, several recent methods have adopted Approximate Bayesian Computation, where supernova samples are forward modelled in parameter space and compared to observed distributions. \citet{Weyant2013} provides an introduction into ABC methods for supernova cosmology in the context of the SDSS-II results \citep{Sako2014} and flat $\Lambda$CDM cosmology, whilst \citet{Jennings2016} demonstrates their \textit{superABC} method on simulated first season Dark Energy Survey samples. In both examples, the supernova simulation package SNANA \citep{Kessler2009a} is used to forward model the data at each point in parameter space.

Simulations provide great flexibility and freedom in how to treat the systematic uncertainties and selection effects associated with supernova surveys. By using forward modelling directly from these simulations, data does not need to be corrected, analytic approximations do not need to be applied, and we are free to incorporate algorithms that simply cannot be expressed in a tractable likelihood such as those found in traditional analyses from Section \ref{sec:traditional}. This freedom comes with a cost -- computation. The classical $\chi^2$ method's most computationally expensive step in a fit is matrix inversion. For ABC methods, we must instead simulate an entire supernova population -- drawing from underlying supernova populations, modelling light curves, applying selection effects, fitting light curves and applying data cuts. This is an intensive process.

One final benefit of ABC methods is that they can move past the traditional treatment of supernovae with summary statistics ($m_B$, $x_1$, and $c$). \citet{Jennings2016} presents two metrics, which are used to measure the distance between the forward modelled population and observed population, and are minimised in fitting. The first metric compares forward modelled summary statistic populations (denoted the `Tripp' metric) and the second utilises the observed supernova light curves themselves, moving past summary statistics. However, we note that evaluation of systematic uncertainty was only performed using the Tripp metric.

\subsection{Bayesian Hierarchical Models}

Sitting between the traditional models simplicity and the complexity of forward modelling lies Bayesian hierarchical models (BHM). Hierarchical models utilise multiple layers of connected parameters, with the layers linked via well defined and physically motivated conditional probabilities. For example, an observation of a parameter from a population will be conditioned on the true value of the parameter, which itself will be conditioned on the population distribution of that parameter. We can thus incorporate different population distributions, and parameter inter-dependence which cannot be found in traditional analyses where uncertainty must be encapsulated in a covariance matrix. Unlike ABC methods, which can model arbitrary probability distributions, BHM methods are generally constrained to representing probabilities using analytic forms.

With the introduction of multiple layers in our model, we can add more flexibility than a traditional analysis whilst still maintaining most of the computational benefits that come from having a tractable likelihood. \citet{Mandel2009, Mandel2011a, Mandel2017} construct a hierarchical model that they apply to supernova light-curve fitting. \citet{March2011} derive a hierarchical model and simplify it by analytically marginalising over nuisance parameters to provide increased flexibility with reduced uncertainty over the traditional method, but do not incorporate bias corrections. \citet{March2014} and \citet{Karpenka2015} improve upon this by incorporating redshift-dependent magnitude corrections from \citet{Perrett2010} to remove bias, and validate on 100 realisations of SNLS-like simulations. The recent BAHAMAS model \citep{Shariff2016} builds on this and reanalyses the JLA dataset \citep[using redshift dependent bias corrections from][]{Betoule2014}, whilst including extra freedom in the correction factors $\alpha$ and $\beta$, finding evidence for redshift dependence on $\beta$. \citet{Ma2016} performed a reanalysis of the JLA dataset within a Bayesian formulation, finding significant differences in $\alpha$ and $\beta$ values from the original analysis from \citet{Betoule2014}. Notably, these methods rely on data that is bias corrected or the methods ignore biases, however the UNITY framework given by \citet{Rubin2015} incorporates selection efficiency analytically in the model, and is applied to the Union 2.1 dataset \citep{Suzuki2012}. The assumption made by the UNITY analysis is that the bias in real data is perfectly described by an analytic function. They validate their  model to be free of significant biases using fits to thirty realisations of supernova datasets that are constructed to mimic the UNITY framework. The well known BEAMS (Bayesian estimation applied to multiple species) methodology from \citet{Kunz2007} has been extended and applied in several works \citep{Hlozek2012}, most lately to include redshift uncertainty for photometric redshift application as zBEAMS \citep{Roberts2017} and to include simulated bias corrections in \citet{Kessler2017}. For the latter case, by inferring biases using Bayesian models, sophisticated corrections can be calculated and then applied to more traditional $\chi^2$ models.

Whilst there are a large number of hierarchical models available, none of them have undergone comprehensive tests using realistic simulations to verify each models' respective bias. Additionally, testing has generally been performed on supernovae simulations with either $\Lambda$CDM cosmology or Flat $\Lambda$CDM cosmology. However, quantifying the biases on $w$CDM cosmology simulations with realistic simulations is becoming critically important as precision supernovae cosmology comes into its own, and focus shifts from determination of $\Omega_m$ to both $\Omega_m$ and $w$.

The flexibility afforded by hierarchical models allows for investigations into different treatments of underlying supernova magnitude, color and stretch populations, host-galaxy corrections, and redshift evolution, each of which will be discussed further in the outline of our model below. Our model is designed to increase the numerical efficiency of prior works whilst incorporating the flexibility of hierarchical models. We reduce our dependence on an assumed scatter model in simulations by not utilising bias-corrections in an effort to provide a valuable cross-check on analysis methodologies which utilise scatter-model-dependant bias corrections.

\section{Challenges in Supernova Cosmology}
\label{sec:challenges}

 The diverse approaches and implementations applied to supernova cosmology are a response to the significant challenges and complications faced by analyses. In this Section, we outline several of the most prominent challenges. 

 Forefront among these challenges is our ignorance of the underlying type Ia intrinsic dispersion. Ideally, analysis of the underlying dispersion would make use of an ensemble of time-series spectroscopy to characterise the diversity of type Ia supernovae. However this data is difficult to obtain, and recent efforts to quantify the dispersion draw inference from photometric measurements. The underlying dispersion model is not a solved problem, and we therefore test against two dispersion models in this work. The first is based on the \citet[][hereafter denoted {\gten}]{Guy2010} scatter model, the second is sourced from \citet[][hereafter denoted {\celeven}]{Chotard2011}. As the SALT2 model does not include full treatment of intrinsic dispersion, each scatter model results in different biases in $m_B$, $x_1$, and $c$ when fitting the SALT2 model to light curve observations, and results in increased uncertainty on the summary statistics that is not encapsulated in the reported covariance $\mathcal{C_\eta}$. These two scatter models are currently assumed to span the possible range of scatter in the underlying supernova population. We have insufficient information to prefer one model over the other, and thus we have to account for both possible scatter models.

The underlying supernova population is further complicated by the presence of outliers. Non-type Ia supernovae often trigger transient follow-up in surveys and can easily be mistaken for type Ia supernovae  and represent outliers from the standardised SN~Ia sample. This contamination is not just a result of non-SN~Ia being observed, but can also arise from host galaxy misidentification causing incorrect redshifts  being assigned to supernovae. Different optimizations to the host-galaxy algorithm can result in misidentification of the host at the 3\% to 9\% level \citep{Gupta2016}, resulting in a broad population of outliers. For spectroscopic surveys, where both supernova type and redshift can be confirmed through the supernova spectra, this outlier population is negligible. However, for photometric surveys, which do not have the spectroscopic confirmation, it is one of the largest challenges; how to model, fit, and correct for contaminants.

Finally, one of the other persistent challenges facing supernova cosmology analyses are the high number of systematics. Because of the rarity of SN~Ia events, datasets are commonly formed from the SN~Ia discoveries of multiple surveys in order to increase the number of supernovae in a dataset. However, each survey introduces additional sources of systematic error, from sources within each survey such as band calibration, to systematics introduced by calibration across surveys. Peculiar velocities, different host environments, and dust extinction represent additional sources of systematic uncertainty which must all be modelled and accounted for.

\section{Our Method}
\label{sec:method}

We construct our hierarchical Bayesian model {\steve} with several goals in mind: creation of a redshift-dependent underlying supernova population,  treatment of an increased number of systematics, and analytic correction of selection effects, including systematic uncertainty on those corrections. We also desire {\steve} more computationally efficient than prior works, such that cosmological results from thousands of supernovae are obtainable in the order of hours, rather than days. As this is closest to the UNITY method from \citet[][hereafter denoted \rubin]{Rubin2015}, we follow a similar model scaffold, and construct the model in the programming language Stan \citep{Carpenter2017, StanDevelopmentTeam2017}. The primary challenge of fitting hierarchical models is their large number of fit parameters, and Stan, which uses automatic differentiation and the no-U-turn Sampler (NUTS, a variant of Hamiltonian Monte Carlo), allows us to efficiently sample high dimensional parameter space.

At the most fundamental level, a supernova cosmology analysis is a mapping from an underlying population onto an observed population, where cosmological parameters are encoded directly in the mapping function. The difficulty arises both in adequately describing the biases in the mapping function, and in adding sufficient, physically motivated flexibility in both the observed and underlying populations whilst not adding \textit{too} much flexibility, such that model fitting becomes pathological due to increasing parameter degeneracies within the model.

\added{In the analysis of this article, the underlying model universe maps to the observed universe as sketched in the BHM of Figure \ref{fig:pgm}. The dependencies that between the model and observations can be tracked following the arrows of the BHM, and a summary of all the conditional probabilities can be found in \ref{sec:model_summary}.

In the following sections, we will describe the model parameters, the mapping functions that connect them to data, the effect of sample selection (in Equation \eqref{eq:main}), and the pathologies that can occur when evaluating the model. Summaries of observables and model parameters are shown in Table 1 for easy reference.}

\subsection{Observed Populations}

\subsubsection{Observables}
\begin{figure}
 	\begin{center}
 		\includegraphics[width=\columnwidth]{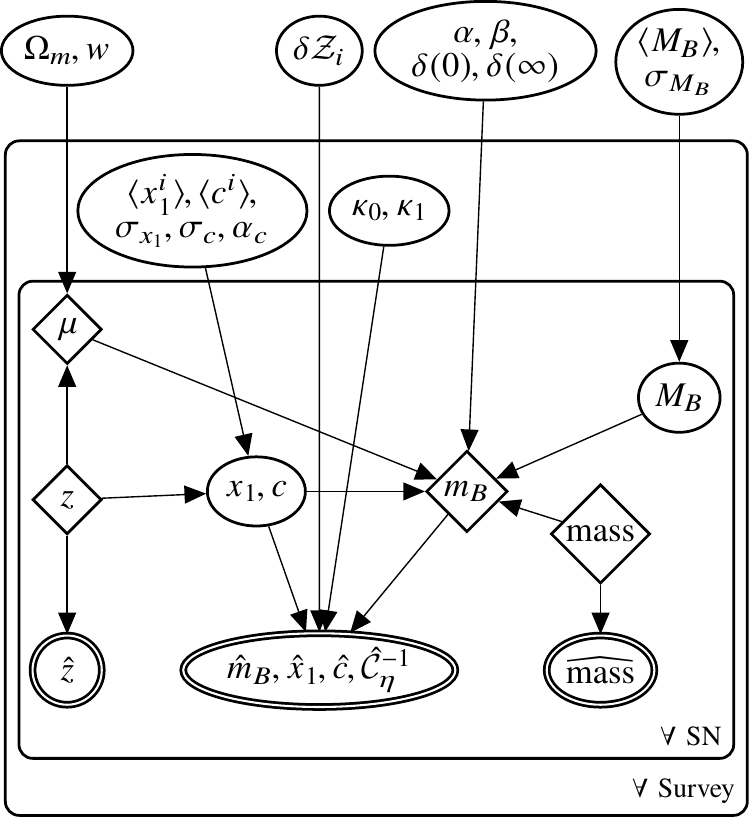}
 	\end{center}
 	\caption{Probabilistic graphical model for our likelihood \added{without selection effects}. Double-lined nodes represent observed variables, diamond nodes represent deterministic variables, single-lined ellipse nodes represent fit variables. The SN box represents observed variables and latent variables for each individual supernova, whilst the survey box represents survey-specific variables, which in general describe the supernova population for the survey and the systematics associated with it. Variables that appear outside both boxes represent top level model parameters. \added{We note that we have shown the model to have latent variables $\lbrace M_B,\,x_1\,c\rbrace$, which uniquely determines $m_B$, given $\mu$ and other parameters. Thus, the two nodes $M_B$ and $m_B$, make up a single layer in our hierarchy, not two layers. In the code implementation, $m_B$ is more efficiently parametrised instead of $M_B$, however the mathematics remains constant regardless of if you parametrise $M_B$ or $m_B$ as one can determine the other. We write out ${\rm mass}$ instead of $m$ to reduce possible confusion with magnitude in the diagram. Finally, as we take redshift measurement $\hat{z}$ and mass probability $\hat{{\rm mass}}$ as exact they are not conditioned on underlying distributions and are top level parameters.}}
 	\label{fig:pgm}
 \end{figure}

Like most of the BHM methods introduced previously, we work from the summary statistics, where each observed supernova has a brightness measurement $\hat{m}_B$ (which is analogous to apparent magnitude), stretch $\hat{x}_1$ and color $\hat{c}$, with uncertainty on those values encoded in the covariance matrix $\cov$. Additionally, each supernova has an observed redshift $\hat{z}$ and a host-galaxy stellar mass associated with it, $\hat{m}$, where the mass measurement is converted into a probability of being above $10^{10}$ solar masses. Our set of observables input into the {\steve} is therefore given as $\lbrace \hat{m}_B, \hat{x}_1, \hat{c}, \hat{z}, \hat{m}, \cov \rbrace$, as shown in the probabilistic graphical model (PGM) in Figure \ref{fig:pgm}.

As we are focused on the spectroscopically confirmed supernovae for this iteration of the method, we assume the observed redshift $\hat{z}$ is the true redshift $z$ such that $P(\hat{z}|z) = \delta(\hat{z} - z)$. Potential sources of redshift error (such as peculiar velocities) are taken into account not via uncertainty on redshift (which is technically challenging to implement \added{as varying redshifts introduce computational complexity in computing the distance modulus integral by reducing the amount of pre-computation that can be utilised}) but instead uncertainty on distance modulus. Similarly, we take the mass probability estimate $\hat{m}$ as correct, and do not model a latent variable to represent uncertainty on the probability estimate. One of the strengths of {\steve} (and the {\rubin} analysis) is that for future data sets where supernovae have been classified photometrically, and we expect some misclassification and misidentification of the host galaxies, these misclassifications can naturally be modelled and taken into account by introducing additional populations that supernovae have a non-zero probability of belonging to.

\begin{deluxetable}{ll}
	\renewcommand{\arraystretch}{0.8}
	\tablecolumns{2}
	\tablewidth{0.99\columnwidth}
	\tablecaption{Model Parameters}
	\tablehead { \colhead{Parameter} & \colhead{Description} }
	\startdata	
	&\textbf{Global Parameters} \\
	$\Omega_m$  & Matter density  \\
	$w$  &  Dark energy equation of state  \\
	$\alpha$ & Stretch standardisation   \\
	$\beta$         &  color standardisation   \\
	$\delta(0)$ & Scale of the mass-magnitude correction\\
	$\delta(\infty)/\delta(0)$ & Redshift-dependence of mass-magnitude correction\\
	$\delta\mathcal{Z}_i$ & Systematics scale\\
	$\langle M_B \rangle$ & Mean absolute magnitude \\
	\\
	& \textbf{Survey Parameters} \\
	$\delta S$ & Selection effect deviation \\
	$\langle x_1^i \rangle$ & Mean stretch nodes\\
	$\langle c^i \rangle$ & Mean color nodes\\
	$\alpha_c$ & Skewness of color population \\
	$\sigma_{M_B}$ & Population magnitude scatter \\
	$\sigma_{x_1}$ & Population stretch scatter \\
	$\sigma_{c}$ & Population color scatter \\
	$\kappa_{0}$ & Extra color dispersion\\
	$\kappa_{1}$ & Redshift-dependence of extra color dispersion\\
	\\
	& \textbf{Supernova Parameters} \\
	$m_B$ & True flux\\
	$x_1$ & True stretch \\
	$c$ & True color \\
	$z$ & True redshift \\	
	$M_B$ & Derived absolute magnitude \\
	$\mu$ & Derived distance modulus \\
	\\
	& \textbf{Input Data}\tablenotemark{a} \\
	$\hat{m}_B$ & Measured flux\\
	$\hat{x}_1$ & Measured stretch \\
	$\hat{c}$ & Measured color \\
	$C$ & Covariance on flux, stretch and color \\
	$\hat{z}$ & Observed redshift \\
	$\hat{m}$ & Observed mass probability \\
	\enddata
	\tablenotetext{a}{Not model parameters but shown for completeness.}
\label{tab:param_summary}
\end{deluxetable}

\subsubsection{Latent Variables for Observables}

The first layer of hierarchy is the set of true (latent) parameters that describe each supernova. In contrast to the observed parameters, the latent parameters are denoted without a hat. For example, $c$ is the true color of the supernova, whilst $\hat{c}$ is the observed color, which, as it has measurement error, is different from $c$.

For the moment, let us consider a single supernova and its classic summary statistics $m_B$, $x_1$, $c$. For convenience, let us define $\eta \equiv \lbrace m_B, x_1, c \rbrace$. A full treatment of the summary statistics would involve determining $p(\hat{y}|\eta)$, where $\hat{y}$ represents the observed light curves fluxes and uncertainties. However, this is computationally prohibitive  as it would require incorporating SALT2 light curve fitting inside our model fitting. Due to this computational expense, we rely on initially fitting the light curve observations to produce a best fit $\hat{\eta}$ along with a $3\times3$ covariance matrix $\mathcal{C_\eta}$ describing the uncertainty on $\hat{\eta}$. Using this simplification, our latent variables are given by
\begin{equation}
p(\hat{\eta}|\eta) \sim \mathcal{N}(\hat{\eta} | \eta, \mathcal{C_\eta}). \label{eq:pop}
\end{equation}
As discussed in Section \ref{sec:challenges},  the SALT2 model does not include full treatment of intrinsic dispersion, and thus this approximation does not encapsulate the full uncertainty introduced from this dispersion.

\subsection{Underlying Population}

\subsubsection{Type Ia population} 
\label{sec:underlying}

Unlike many previous formalisms which utilise $M_B$ as a singular number and model magnitude scatter on the apparent magnitude $m_B$, we incorporate this scatter into the underlying rest-frame population by having a population in absolute magnitude space. \added{This is mathematically equivalent, however allows us to model the underlying population and intrinsic scatter distinctly.} To denote this difference, we refer to the mean of our absolute magnitude population with $\langle M_B \rangle$.

In addition to absolute magnitude, the underlying supernova population is also characterised by distributions in color and stretch. We follow the prior work of {\rubin} and model the color population as an independent redshift-dependent skew normal distribution for each survey. For the stretch population, we adopt a redshift-dependent normal distribution, and magnitude dispersion is modelled as a normal distribution. We also tested a skew-normal approach for these parameters, reverting to the normal distributions as they are computationally easier to evaluate and we found no reduction in cosmological bias with the skew-normal distributions for stretch and magnitude. Following {\rubin} we allow the mean color and stretch to vary over redshift, anchoring four equally spaced redshift nodes spanning the redshift range of each survey, linearly interpolating between the nodes to determine the mean stretch and color for a given redshift. These nodes are represented as $\langle x_1^i \rangle$ and $\langle c^i \rangle$. Both the color and stretch means are modelled with normal priors. Initial versions of our model adopted a fully covariant multivariate skew normal (with skewness set to zero only for the magnitude component) to capture correlations between $m_B$ and $c$, however pathological fitting complications required us to simplify our treatment. We parametrise the color skewness $\alpha_c$ by sampling $\delta_c = \alpha_c / \sqrt{1 + \alpha_c^2}$ which itself is given a uniform prior $\mathcal{U}(0,0.98)$ that allows $\alpha_c$ to span positive values in the realm of physical plausibility as determined from constraints in \citet{Scolnic2016}. We sample $\delta_c$ in log space for efficiency in sampling close to zero. The width of the population, represented by the vector $\lbrace \sigma_{M_B}, \sigma_{x_1}, \sigma_c \rbrace$ is subject to Cauchy priors \added{with mean 0 and width 1, following recommendations from the Stan user guide}.

The only constant between survey populations is the absolute magnitude $\langle M_B \rangle$. We model the colour skewness and the redshift-dependent means and width of the colour and skew distributions individually for each survey. The probability for a supernova to have true values $M_B$, $x_1$, $c$ given the underlying population is thus given as
\begin{eqnarray}
P(M_B, x_1, c, z | \theta) &=& \mathcal{N}(M_B|\langle M_B \rangle, \sigma_{M_B}) \, \mathcal{N}(x_1 | \langle x_1(z) \rangle, \sigma_{x_1})\notag\\ &\quad\quad& \mathcal{N}^{\rm{skew}}(c| \langle c(z) \rangle, \sigma_c, \alpha_c) \label{eq:l1},
\end{eqnarray}
where $\theta = \lbrace \langle M_B \rangle, \langle x_1(z) \rangle, \langle c(z) \rangle, \sigma_{m_B}, \sigma_{x_1}, \sigma_c, \alpha_c \rbrace$ for legibility.

\subsubsection{Outlier populations}

\added{For the spectroscopic paper we do not consider outlier populations, however we ensure that our model has flexibility for such populations for future use with photometrically classified surveys. We thus include a simplistic outlier population model.} We follow {\rubin} \citep[and therefore ][]{Kunz2007} by implementing a Gaussian mixture model, where an additional observable of the SN~Ia probability would be needed in order to inform the weights of the mixture model. For surveys with low rates of contamination, it is not possible to fit a contamination population, and the mean of the outlier population has been fixed to the SN~Ia population in prior works. However, with the increased number of contaminants expected in the DES photometric sample, we seek a more physically motivated outlier population.  We find that an acceptable parametrisation is to model the outlier population with a mean absolute magnitude of $\langle M_B^{\rm outl} \rangle = \langle M_B \rangle + \delta_{M_B}^{\rm outl}$, where $\delta_{M_B}^{\rm outl}$ is constrained to be positive, or even to be greater than a small positive number to reduce degeneracy between the two populations.  We note that this represents the mean brightness of outliers, and so outliers could both be brighter and dimmer than the mean SN~Ia absolute magnitude. We set the population width to $\sigma^{\rm outl} = 1$ in $M_B$, $x_1$ and $c$ in our tests. The probability of each supernova falling into either population is determined by the observed type Ia probability $\hat{p}$. For the spectroscopic survey, we set this to unity, and thus it is not included in Figure \ref{fig:pgm} or Table \ref{tab:param_summary}. For the photometric proof-of-concept we provide an accurate probability estimate. Further investigation on the effect of inaccurate estimates will be left for future improvements during the analysis of the DES photometric sample.

\subsection{Correcting biased summary statistics}

With the fitted summary statistics $\hat{\eta}$ being biased and their uncertainty under-reported, we face a significant challenge utilising these statistics naively in supernova cosmology. We must either correct the observables to remove the biases introduced by the intrinsic dispersion of the underlying population, or incorporate this dispersion into our model. We should also avoid assuming a specific dispersion model -- either the {\gten} or {\celeven} model, or utilise the results of computing the bias from both models.

We model the extra dispersion only in color, and do so by adding independent uncertainty on the color observation $\hat{c}$. We note that extra dispersion in magnitude $\hat{m}_B$ (from coherent scatter) is absorbed completely by the width of the underlying magnitude population (discussed in Section \ref{sec:underlying}) without introducing cosmological bias, which is not true of the color term, hence the requirement for modelling additional color dispersion. Tests on incorporating extra dispersion on stretch as well show that stretch is less biased than color, and causes negligible bias in cosmology.

As shown in \citep{Kessler2013}, the extra color dispersion shows heavy redshift dependence, increasing with redshift. This is an artifact of different filters, however as we may be subject to similar effects in our observational data, we decide to incorporate redshift dependence in our extra uncertainty. We thus add $\kappa_0 + \kappa_1 z$ to our observed color uncertainty (in quadrature). The $\kappa$ parameters are highly degenerate with the width of the intrinsic color population $\sigma_c$. We subject them to Cauchy priors centered on zero and with width $0.05$, where $\kappa$ is bounded between $0$ and $0.05$. We pick this maximum value to allow extra dispersion without completely subsuming the intrinsic population widths due to the severe degeneracy, where this maximum value easily encapsulates the determined dispersion according to the results of \citet{Kessler2013}. As such, our combined covariance on the observation $\hat{\eta}$ is given by $\mathcal{C}_{\rm tot} = \mathcal{C_\eta} + {\rm DiagMatrix}\left[0, 0, (\kappa_0 + \kappa_1 z)^2\right]$. 

Fully covariant extra dispersion on $\lbrace m_B, x_1, c \rbrace$ (rather than just dispersion on $c$) was also tested, by modelling the dispersion as a multivariate Gaussian, but it showed negligible improvement in recovering unbiased cosmology over just color dispersion, and was far more computationally inefficient. We note here that we model dispersion in magnitude, but this is done at the level of underlying populations, not observed populations. This magnitude dispersion is modelled with redshift independence.

\subsection{Mapping function}

\subsubsection{Cosmology}

We formulate our model with three different cosmological parameterisations; Flat $\Lambda$CDM, Flat $w$CDM, and standard $\Lambda$CDM \added{ (without a flatness prior)}. $\Omega_m$ is given the prior $\mathcal{U}(0.05, 0.99)$, $\Omega_\Lambda$ was treated with $\mathcal{U}(0, 1.5)$ and the equation of state $w$ was similarly set to a flat prior $\mathcal{U}(-0.4, -2.0)$. For calculating the distance modulus, we fix $H_0 = 70 \kmsmpc $. If the Hubble constant has a different value, the absolute magnitude is $M_B + 5\log(H_0/70 \kmsmpc )$ with the other cosmological parameters unaffected.

\subsubsection{Supernova Standardisation Parameters}

With increasingly large datasets and more nuanced analyses, the choice of how to handle $\alpha$ and $\beta$ becomes an important consideration when constructing a model. {\rubin} employs a broken linear relationship for both color and stretch, where different values of $\alpha$ and $\beta$ are adopted depending on whether $x_1$ and $c$ are positive or negative (although the cut could be placed at a location other than 0). \citet{Shariff2016} instead model $\beta$ as redshift-dependent, testing two phenomenological models; $\beta(z) = \beta_0 + \beta_1 z$ and a second model which effects a rapid but smooth change in $\beta$ at a turnover redshift $z_t$.

We tested two models with varying $\beta$ against simulated supernova sets; $\beta(c) = \beta_0 + \beta_1 c$ and $\beta(z) = \beta_0 + \beta_1 z$. See Section \ref{sec:simdes} for details on simulation generation. We found for both models that non-zero values for $\beta_1$ are preferred even with constant $\beta$ used in simulation, due to severe degeneracy with selection effects. This degeneracy resulted in a significant bias in recovered cosmology. Due to the recovery of non-zero $\beta_1$, we continue to adopt the constant $\alpha$ and $\beta$ found in traditional analyses. As such, our calculation of distance modulus $\mu$ mirrors that found in Equation \eqref{eq:mu0}.

\subsubsection{Host Galaxy Environment}
\label{sec:hostgal}

There are numerous results showing statistically significant correlations between host-galaxy environment and supernova properties \citep{Kelly2010, Lampeitl2010, Sullivan2010, DAndrea2011, Gupta2011, Johansson2013, Rigault2013}. The latest sample of over 1300 spectroscopically confirmed type Ia supernovae show $>5\sigma$ evidence for correlation between host mass and luminosity \citep{Uddin2017}. The traditional correction, as employed in analyses such as \citet{Suzuki2012} and \citet{Betoule2014}, invokes a step function such that $\Delta M = \gamma \mathcal{H}(\log(M) - 10))$, where $\mathcal{H}$ is the Heaviside step function, $M$ is the galaxy mass in solar masses and $\gamma$ represents the size of the magnitude step. The scale of this step function varies from analysis to analysis, and is treated as a fit parameter. In this work we adopt the model used in {\rubin}, which follows the work from \citet{Rigault2013}, such that we introduce two parameters to incorporate a redshift-dependent host galaxy mass correction:
\begin{equation}
\Delta M = \delta(0) \left[ \frac{1.9\left(1 - \frac{\delta(0)}{\delta(\infty)}\right)  }{0.9 + 10^{0.95z}} + \frac{\delta(0)}{\delta(\infty)}\right], \label{eq:mass}
\end{equation}
where $\delta(0)$ represents the correction at redshift zero, and $\delta(\infty)$ a parameter allowing the behaviour to change with increasing redshift. We take flat priors on $\delta(0)$ and $\delta(0)/\delta(\infty)$. \added{Finally, we assume that the observed mass probability $\hat{m}$ supplied to the model is perfectly determined, and thus set $P(\hat{m}|m) = \delta(\hat{m} - m)$.}

\subsubsection{Uncertainty Propagation}
\label{sec:systreat}
The chief difficulty with including systematic uncertainties in supernova analyses is that they have difficult-to-model effects on the output observations. As such, the traditional treatment for systematics is to compute their effect on the supernova summary statistics -- computing the numerical derivatives $\frac{{\rm d} \hat{m}_B}{{\rm d} \Z_i}$, $\frac{{\rm d} \hat{x}_1}{{\rm d} \Z_i}$, $\frac{{\rm d} \hat{c}}{{\rm d} \Z_i}$ \added{ for each supernova light curve fit}, where $\Z_i$ represents the $i$\textsuperscript{th} systematic.

Assuming that the gradients can be linearly extrapolated -- which is a reasonable approximation for modern surveys with high quality control of systematics -- we can incorporate into our model a deviation from the observed values by constructing a $(3 \times N_{\rm sys})$ matrix containing the numerical derivatives for the $N_{\rm sys}$ systematics and multiplying it with the row vector containing the offset for each systematic. By scaling the gradient matrix to represent the shift over $1\sigma$ of systematic uncertainty, we can simply enforce a unit normal prior on the systematic row vector to increase computational efficiency.

This method of creating a secondary covariance matrix using partial derivatives is used throughout the traditional and BHM analyses. For each survey and band, we have two systematics --- the calibration uncertainty and the filter wavelength uncertainty. We include these in our approach, in addition to including HST Calspec calibration uncertainty, ten SALT2 model systematic uncertainties, a dust systematic, a global redshift bias systematic, and also the systematic peculiar velocity uncertainty. A comprehensive explanation of all systematics is given in \citet{Brout18SYS}; see Table~4 for details. This gives thirteen global systematics shared by \added{ all surveys that apply globally to all supernova summary statistics}, plus two systematics per band in each survey. \added{Systematics with known correlations are shifted together to produce covariant deviations, and we thus assume that the numerical derivatives input into our model represent independent systematics. Full details can be found in \citet{Brout18SYS}.} With $\eta \equiv \lbrace m_B, x_1, c \rbrace$, our initial conditional likelihood for our observed summary statistics shown in Equation \eqref{eq:pop} becomes
\begin{equation}
P\left(\hat{\eta}, \frac{\partial \hat{\eta}}{\partial \Z_i} | \eta, \delta \Z_i, C_\eta\right) = \mathcal{N}\left(\hat{\eta} + \delta \Z_i \frac{\partial \hat{\eta}}{\partial \Z_i}|\eta,\cov\right). \label{eq:l3}
\end{equation}

\subsubsection{Selection Effects}
\label{sec:selection}

One large difference between traditional methods and BHM methods is that we treat selection effects by incorporating selection efficiency into our model, rather than relying on simulation-driven data corrections. We describe the probability that the \added{ possible} events we observe are drawn from the distribution predicted by the underlying theoretical model \textit{and} that those events, given they happened, are observed and pass cuts.  To make this extra conditional explicit, we can write the likelihood of the data given an underlying model, $\theta$, \textit{and} that the data are included in our sample, denoted by $S$, as
\begin{align}
\mathcal{L}(\theta; {\rm data}) &= P({\rm data} | \theta, S). \label{eq:like}
\end{align}
As our model so far describes components of a basic likelihood $P({\rm data}|\theta)$, and we wish to formulate a function $P(S|{\rm data},\theta)$ that describes the chance of an event being successfully observed, we rearrange the likelihood in terms of those functions and find
\begin{align}
\mathcal{L}(\theta; {\rm data}) &= \frac{P(S|{\rm data},\theta) P({\rm data}|\theta)}{\int P(S | D, \theta) P(D|\theta)\, dD}, 
\end{align}
where the denominator represents an integral over all potential data $D$, \added{and $\theta$ represents top-level parameters. In the case that our selection effects are best characterised by latent variables \textit{instead of} data, we can add them to our formulation and our likelihood becomes
\begin{eqnarray}
\mathcal{L}(\theta; {\rm data})&=& \frac{\int P(S|L,\theta) P({\rm data}|L) P(L|\theta)\, dL}{\iint P(S|L,\theta) P(L|\theta) dL},\label{eq:main}
\end{eqnarray}
where $L$ represents our latent parameters.}
This is derived in Appendix \ref{app:selection1}. \added{ To evaluate the effect of our selection effects, we need to evaluate both the selection effect terms in the numerator and the integral in the denominator. The numerator represents the probability we caught the supernova and it was selected into the cosmology sample. The integral represents our global selection efficiency at a location in parameter space, rather than the probability of our data being selected into our sample.} As $\theta$ represents the vector of all \added{top-level model parameters, and $L$ represents a vector of all latent parameters, this is not a trivial integral.} Techniques to approximate this integral, such as Monte-Carlo integration or high-dimensional Gaussian processes failed to give tractable posterior surfaces that could be sampled efficiently by Hamiltonian Monte-Carlo, and post-fitting importance sampling failed due to high-dimensionality (a brief dismissal of many months of struggle). We therefore simplify the integral and approximate the selection effects from their full expression in all of $\theta$-space, to apparent magnitude and redshift space independently (not dependent on $x_1$ or $c$), such that the denominator of equation \eqref{eq:main}, denoted now $d$ for simplicity, is given as
\begin{equation}
d = \int  \left[ \int P(S|m_B) P(m_B | z, \theta)\, d m_B \right] P(S|z) P(z|\theta)\, dz, \label{eq:w1}
\end{equation}
where $P(m_B | z, \theta)$ can be expressed by translating the underlying $M_B$, $x_1$, and $c$ population to $m_B$ given cosmological parameters. A full derivation of this can be found in Appendix \ref{app:selection2}.

We now apply two further approximations similar to those made in {\rubin} -- that the redshift distribution of the observed supernovae reasonably well samples the $P(S|z)P(z|\theta)$ distribution, and that the survey color and stretch populations can be treated as Gaussian for the purposes of evaluating $P(m_B | z, \theta)$. We found that discarding the color population skewness entirely resulted in highly biased population recovery (see Figure \ref{fig:simple_w_super} to see the populations), and so we instead characterise the skew normal color distribution with a Gaussian that follows the mean and variance of a skew normal; with mean given by $\langle c(z) \rangle + \sqrt{\frac{2}{\pi}} \sigma_c \delta_c$ and variance $\sigma_c^2(1 - 2\delta_c^2/\pi)$. This shifted Gaussian approximation for color completely removes the unintended bias when simply discarding skewness. This shift was not required for the stretch population, and so was left out for the stretch population for numerical reasons. The impact of this approximation on the calculated efficiency is shown in Figure \ref{fig:shift}, and more detail on this shift and resulting population recovery can be found in Appendix \ref{app:approx}. 

\begin{figure}
	\begin{center}
		\includegraphics[width=\columnwidth]{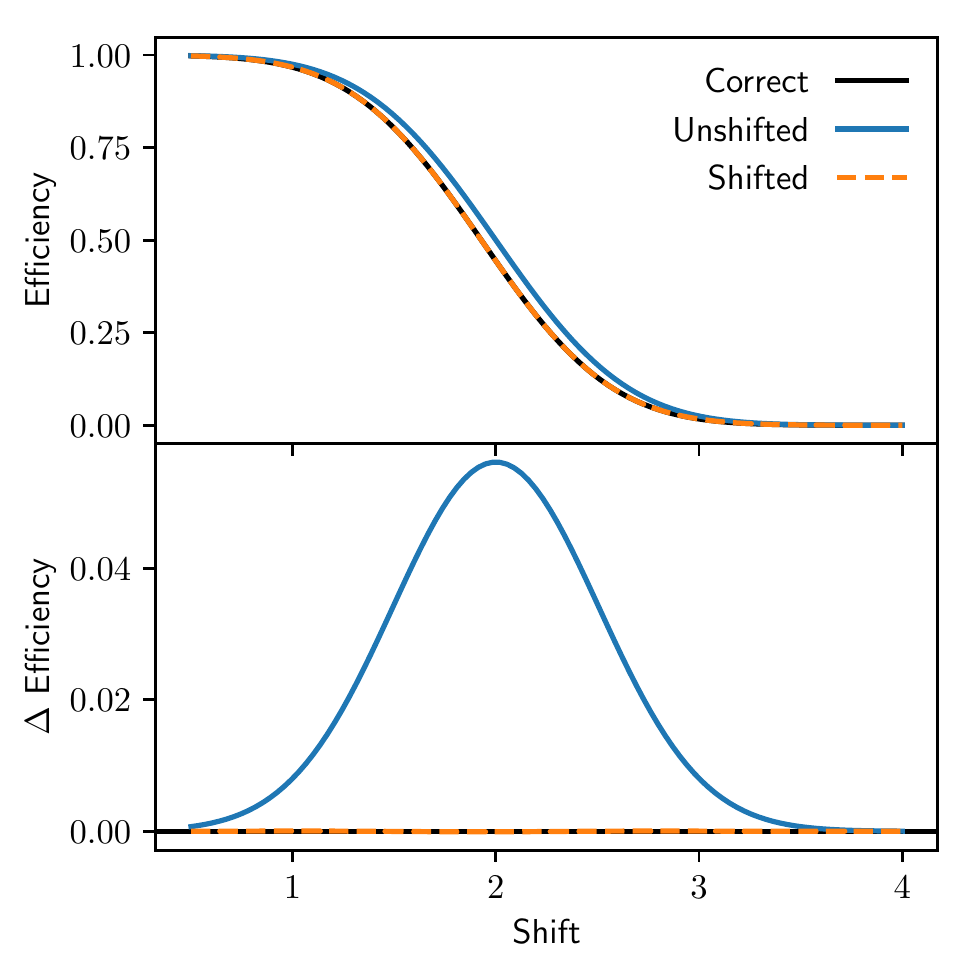}
	\end{center}
	\caption{Testing the correctness of our normal approximation to the skewed color distribution. The `correct' line (shown in black) represents the exact integral $w = \int P(S|x) P(x) dx$ where $P(S|x)$ is an error function (following our high-redshift surveys) and $P(x) = \mathcal{N}^{\rm Skew}(x, 0.1, 2)$, calculated numerically. The $x$-axis is analogous to $m_B$ is cosmological context. As expected, all efficiencies drop towards zero as shift increases (as objects get fainter). The unshifted normal approximation shows significant discrepancy in the calculated efficiency as it transitions from 1 to 0, whilst the shifted normal approximation shows negligible error to the correct solution. From these plots, further refinement of the normal approximation (such as including kurtosis or higher powers) are unnecessary.}
	\label{fig:shift}
\end{figure}

The population $P(m_B | z, \theta)$ becomes $\mathcal{N}(m_B|m_B^*(z), \sigma^*_{m_B})$, where
\begin{eqnarray}
m_B^*(z) &=& \langle M_B \rangle + \mu(z) - \alpha \langle x_1(z) \rangle + \beta \langle c(z) \rangle \\
\sigma^{*2}_{m_B} &=& \sigma_{M_B}^2 + (\alpha \sigma_{x_1})^2 + (\beta \sigma_c)^2.
\end{eqnarray}
What then remains is determining the functional form of $P(S|m_B)$. For the treatment of most surveys, we find that the error function, which smoothly transitions from some constant efficiency down to zero, is sufficient. Formally, this gives
\begin{equation}
P(S|m_B) = \Phi^c(m_B | \mu_{\rm CDF}, \sigma_{\rm CDF}), \label{eq:selcdf}
\end{equation}
where $\Phi^c$ is the complementary cumulative distribution function and $\mu_{\rm CDF}$ and $\sigma_{\rm CDF}$ specify the selection function. The appropriateness of an error function has been found by many past surveys \citep{Dilday2008, Barbary2010, Perrett2012, Graur2013, Rodney2014}. However, for surveys which suffer from saturation and thus rejection of low-redshift supernovae, or for groups of surveys treated together (as is common to do with low-redshift surveys), we find that a skew normal is a better analytic form, taking the form
\begin{equation}
P(S|m_B) = \mathcal{N}^{{\rm Skew}}(m_B | \mu_{\rm Skew}, \sigma_{\rm Skew}, \alpha_{\rm Skew}). \label{eq:selnorm}
\end{equation}

The selection functions are fit to apparent magnitude efficiency ratios calculated from SNANA simulations, by taking the ratio of supernovae that are observed and passed cuts over the total number of supernovae generated in that apparent magnitude bin. That is, we calculate the probability we would include a particular supernova in our sample, divided by the number of such supernovae in our simulated fields. To take into account the uncertainty introduced by the imperfection of our analytic fit to the efficiency ratio, uncertainty was uniformly added in quadrature to the efficiency ratio data from our simulations until the reduced $\chi^2$ of the analytic fit reached one, allowing us to extract an uncertainty covariance matrix for our analytic fits to either the error function or the skew normal. This is mathematically identical to fitting the efficiency ratio with a second `intrinsic dispersion' parameter which adds uncertainty to the efficiency ratio data points.

We thus include into our model parametrised selection effects by including the covariance matrix of selection effect uncertainty. Formally, we include deviations from the determined mean selection function parameters with parameter vector $\delta S$, and apply a normal prior on this parameter as per the determined uncertainty covariance matrix.  Whilst this uncertainty encapsulates the potential error from the simulations not matching the analytic approximations, it does not cover potential variations of the selection function at the top level --- varying cosmology or spectroscopic efficiency. Tests with changing the intrinsic scatter model used in the selection efficiency simulations show that the uncertainty introduced is negligible.

\begin{figure}
	\begin{center}
		\includegraphics[width=\columnwidth]{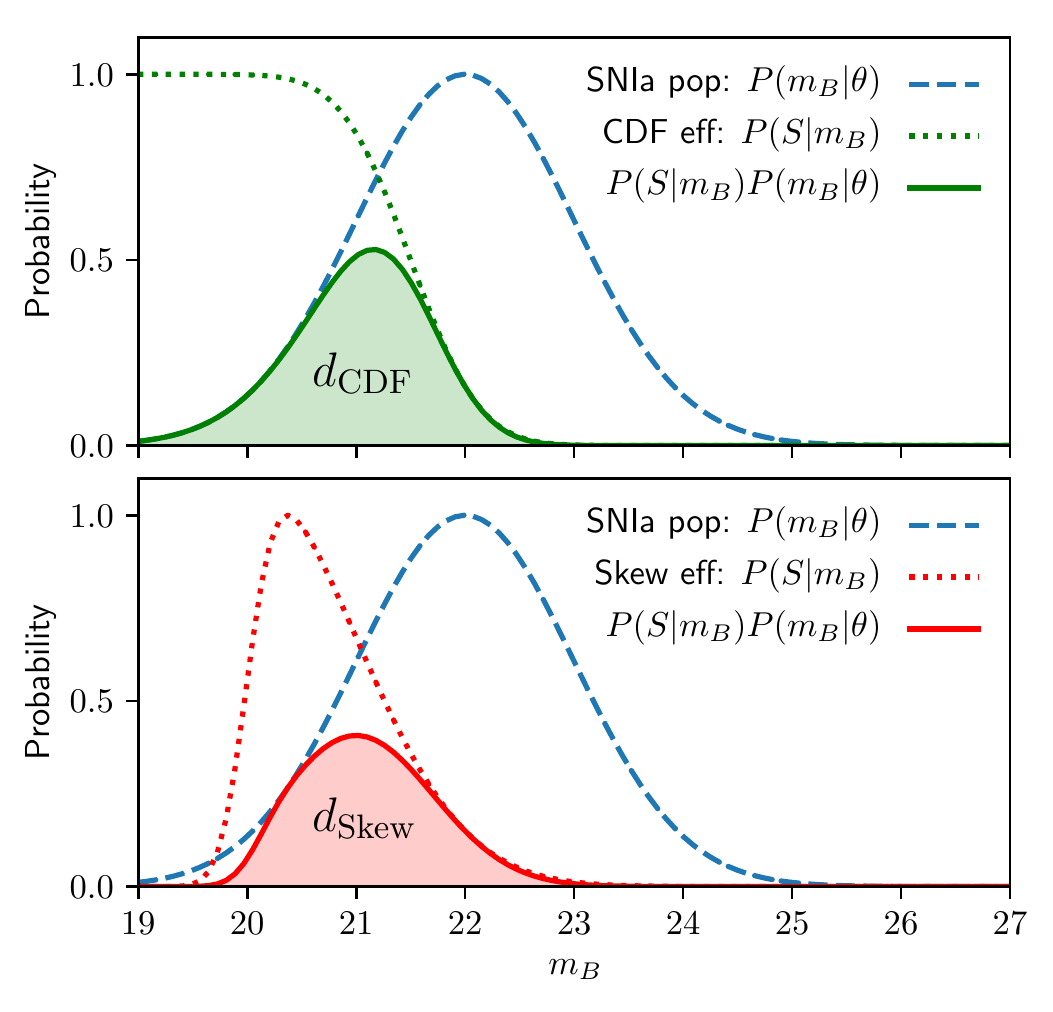}
	\end{center}
	\caption{The efficiency for supernova discovery at an arbitrary redshift. Shown in both panels in dashed blue is the SN~Ia population distribution, which takes the form of a normal distribution. The top panel shows a CDF based survey efficiency (green dotted line), whilst the bottom panel shows a skew normal based survey efficiency (red dotted line), as functions of apparent magnitude. The survey efficiency, given the SN~Ia population, is shown as a solid line in both panels, and the probability of observing a SN~Ia is found by integrating over the population detection efficiency as described in equation \eqref{eq:w1}, and has been shown by shading the area integrated. This area is what is analytically given by equations \eqref{eq:seldes} and \eqref{eq:sellow-z}.}
	\label{fig:efficiency}
\end{figure}

With the well-sampled redshift approximation we can remove the redshift integral in Eq \eqref{eq:w1} and replace it with a correction for each observed supernova. For the error function (denoted with the subscript `CDF') and skew normal selection functions respectively (denoted with a subscript `Skew'), the correction \textit{per SN~Ia} becomes
\begin{eqnarray}
d_{\rm CDF} &=& \Phi^c\left(  \frac{m^*_B - \mu_{\rm CDF}}{\sqrt{\sigma^{*2}_{m_B} + \sigma_{\rm CDF}^2}}  \right) \label{eq:seldes}\\
d_{\rm Skew} &=& 2\mathcal{N}\left( \frac{m_B^* - \mu_{\rm Skew}}{\sqrt{\sigma^{*2}_{m_B} + \sigma_{\rm Skew}^2}}\right) \times \notag\\
&\quad\quad& \Phi\left(  \frac{{\rm sign}(\alpha_{\rm Skew})(m_B^* - \mu_{\rm Skew})}{\frac{\sigma_{m_B}^{*2} + \sigma^2_{\rm Skew}}{\sigma^2_{\rm Skew}} \sqrt{\frac{\sigma^2_{\rm Skew}}{\alpha^2_{\rm Skew}} + \frac{\sigma_{m_B}^{*2} \sigma^2_{\rm Skew}}{\sigma_{m_B}^{*2} + \sigma^2_{\rm Skew}}} }\right), \label{eq:sellow-z}
\end{eqnarray}
and is incorporated into our likelihood. \added{ Note that the above efficiencies utilise the common form of the normal distribution rather than the conditional probability notation found previously in this work.} This is illustrated in Figure \ref{fig:efficiency}. Our corrections for the DES spectroscopic data utilise the CDF functional form, with the combined low redshift surveys being modelled with the skew normal efficiency. Further details on this choice are given in Section \ref{sec:simdes}.

\added{
\subsection{Model Summary}
\label{sec:model_summary}

Having laid out each individual aspect of the model, the relationships between variables and our treatment of uncertainty, here we summarise the relationships in our model mathematically. In this summary of $P({\rm data} | \theta)$, we leave out sample selection for simplicity. Referring to  equation \eqref{eq:main}, the relationships with $P({\rm data} | \theta)$ are as follows:

\begin{align}
&P({\rm data}|\theta) \propto \notag \\
&\int\, d\mu\, dm_B\, dx_1\, dc\, dz, dm \  P(\hat{z}|z) P(\hat{m}|m)  \notag \\
& P(\hat{m}_B, \hat{x}_1, \hat{c}, \hat{\mathcal{C}}_\eta^{-1} | m_B, x_1, c, \delta \mathcal{Z}_i, \kappa_0, \kappa_1) \notag \\
& P(x_1, c |  \langle x_1^i \rangle, \langle c^i \rangle, \sigma_{x_1}, \sigma_c, z) \notag\\
& P(m_B | \langle M_B \rangle, \mu, \delta(0), \delta(\infty), \sigma_{M_B}, \sigma_{x_1}, \sigma_c, \alpha, \beta, z, m) \notag \\
& P(\mu |z, \Omega_m, w).
\end{align}

The denominator of equation \eqref{eq:main} is then given by either equation \eqref{eq:seldes} or \eqref{eq:sellow-z} depending on the survey, and similarly $P(S|{\rm data}, \theta)$ is given by equation \eqref{eq:selnorm} or \eqref{eq:selcdf} respectively. Combining all of these gives us our full model likelihood with selection effects accounted for.

}

\section{Model Verification}
\label{sec:verification}

In order to verify our model we run it through stringent tests. First, we validate on toy models, verifying that we recover accurate cosmology when generating toy supernovae data constructed to satisfy the assumptions of the BHM construction. We then validate our model on SNANA simulations based on a collection of low redshift surveys and the DES three-year spectroscopic sample, termed the DES-SN3YR sample

\subsection{Applied to Toy Spectroscopic Data}
\label{sec:toy}

\begin{figure}
	\begin{center}
		\includegraphics[width=\columnwidth]{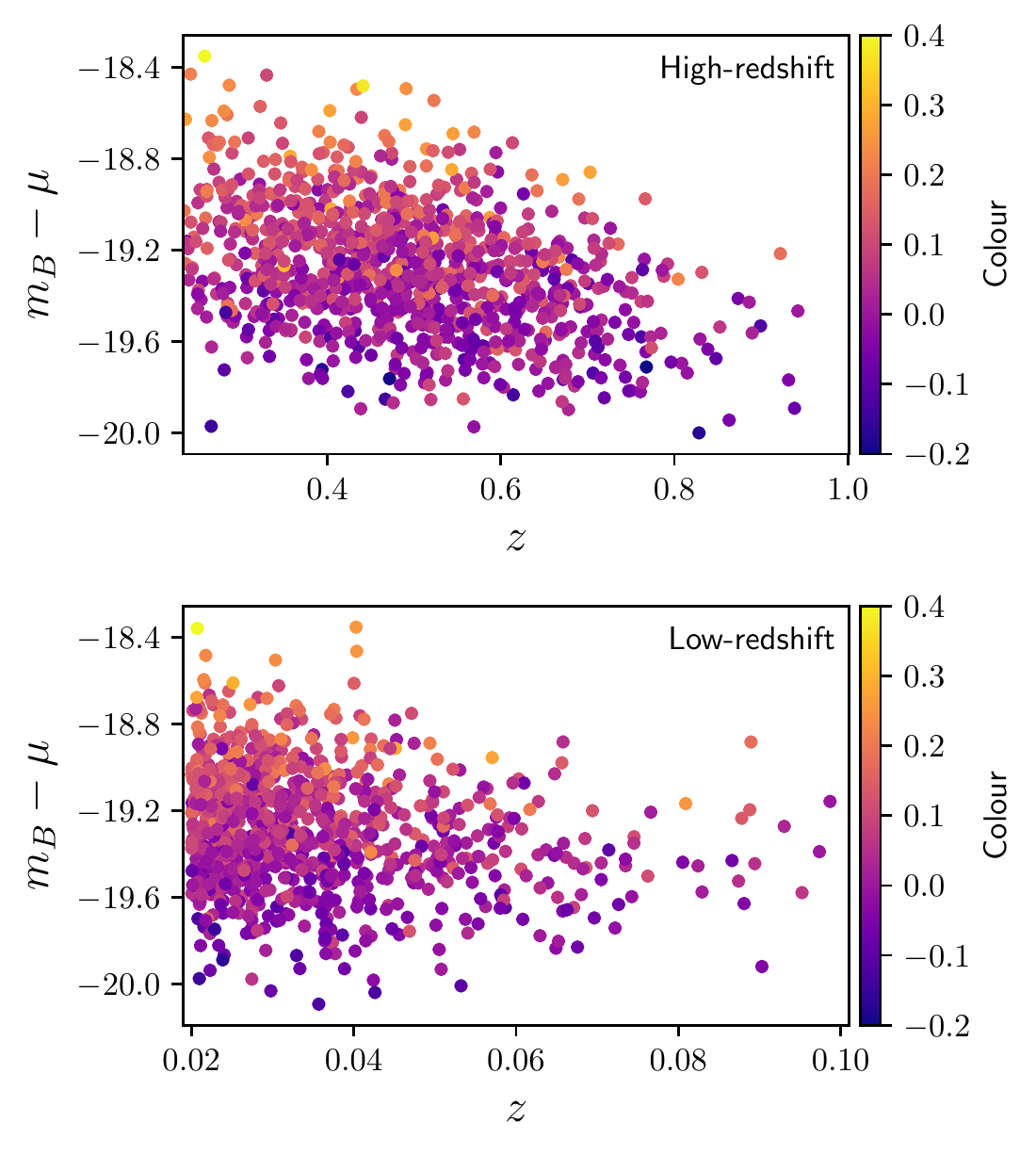}
	\end{center}
	\caption{Population distributions shown in redshift and uncorrected absolute magnitude $m_B - \mu$ for 1000 supernovae in both high-redshift and low-redshift surveys. Selection effects are visible in both samples, where red supernovae are often cut as redshift increases, creating a skewed color population. The color of the data points is representative of the supernova color itself, a negative color value showing bluer supernovae, with positive color values representing redder supernovae.}
	\label{fig:simple_pop}
\end{figure}
\begin{figure}
	\begin{center}
		\includegraphics[width=\columnwidth]{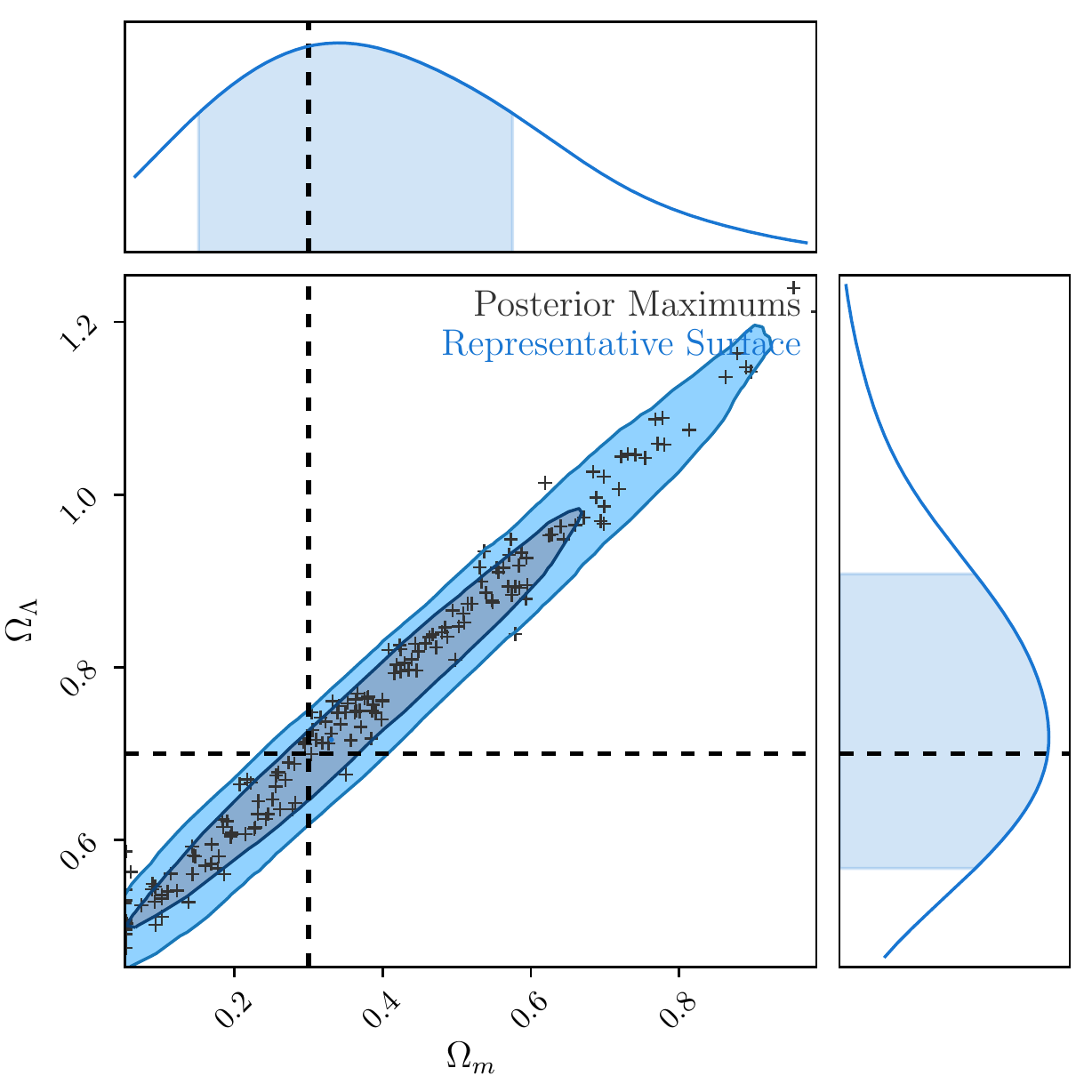}
	\end{center}
	\caption{Maximal posterior points for 100 realisations of supernova data with the Flat $\Lambda$CDM model, with a representative contour from a single data realisation shown for context. Even a large supernova sample when treated robustly is insufficient to provide tight constraints on either $\Omega_m$ or $\Omega_\Lambda$ separately due to the severe degeneracy between the parameters.}
	\label{fig:simple_ol}
\end{figure}
\begin{figure}
	\begin{center}
		\includegraphics[width=\columnwidth]{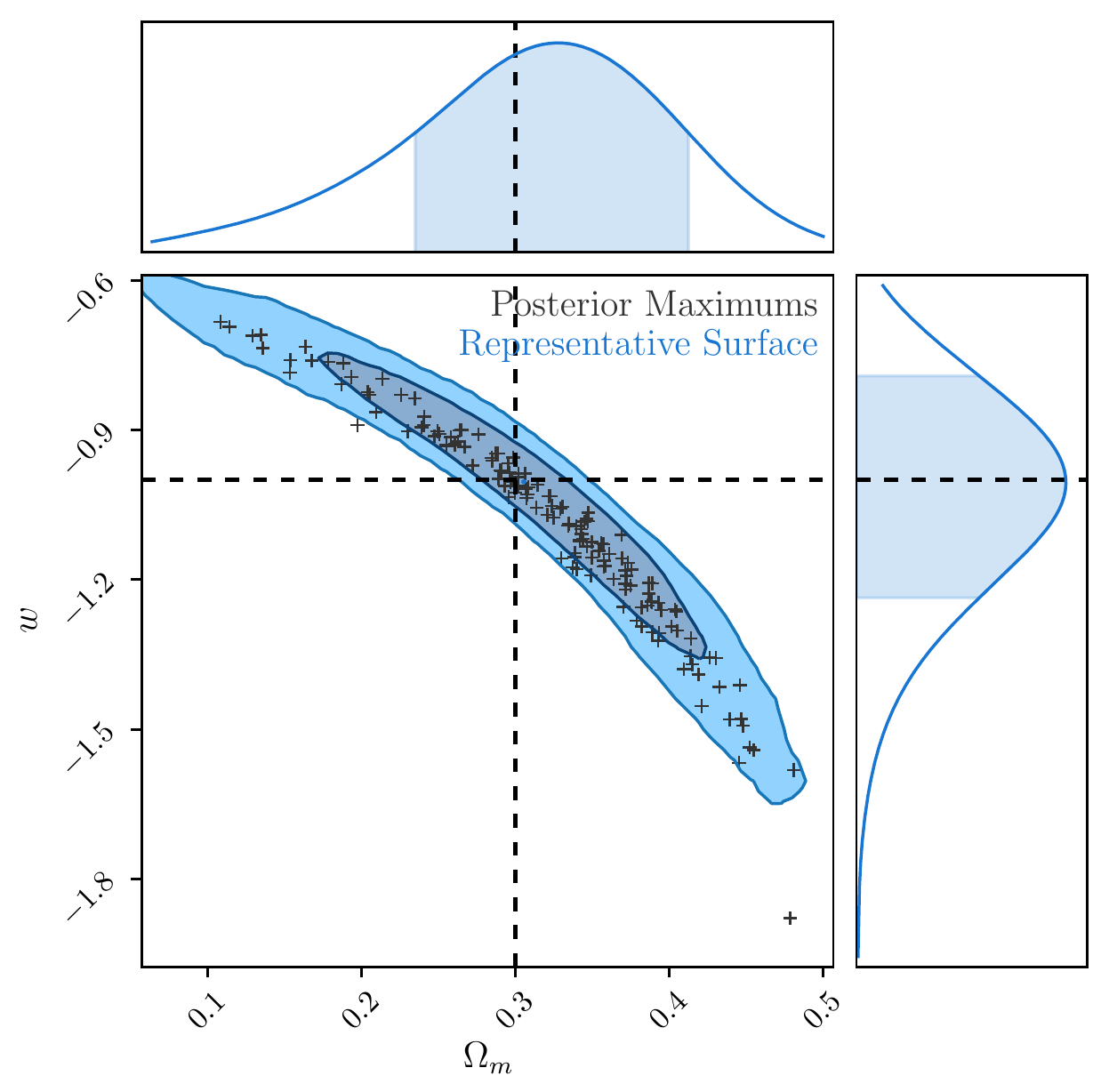}
	\end{center}
	\caption{Maximal posterior points for 100 realisations of supernova data with the Flat $w$CDM model, with a representative contour from a single data realisation shown for context. The well known banana shaped contour is recovered, with the marginalised distributions in $\Omega_m$ and $w$ \added{ shown to recover input cosmology. For contours that are non-Gaussian due to the curved degeneracy between $\Omega_m$ and $w$, the marginalised distributions can provide misleading statistics where maximum marginalised distribution can disagree with the maximum likelihood in multiple dimensions. For our contours, the non-Gaussianity is small and the marginalised distributions still provide a valuable metric.} The recovered posteior maximums show the same degeneracy direction as the representative surface, and scatter around the truth values input into the simulation, which are shown in dashed lines.}
	\label{fig:simple_w}
\end{figure}

\begin{deluxetable}{lcc}
	\tablecolumns{3}
	\tablecaption{Cosmological Parameters for toy supernova data}
	\tablehead {
		\colhead {Model}              &
		\colhead {$\Omega_m\, \langle\mu\rangle,\langle\sigma\rangle$(scatter)}  &
		\colhead{$w\, \langle\mu\rangle,\langle\sigma\rangle$(scatter)}
	}
	
	\startdata
		Flat $\Lambda$CDM & $0.301,\, 0.015\,(0.012)$ & -- \\ 
		Flat $w$CDM & -- & $-1.00,\, 0.042\,(0.030)$ \\ 
	\enddata
	\label{tab:simple_model}
	\tablecomments{Cosmological parameters determined from the surfaces of 100 fits to independent realisations of toy supernova data. As described in the main text, each dataset comprised 1000 low-redshift supernovae and 1000 high-redshift supernovae. For each chain, we record the mean and standard deviation, and then show the average mean and average standard deviation in the table. The scatter introduced by simulation variance (the standard deviation of the 100 mean parameter values) is shown in brackets. Model bias would appear as shifts away from the simulation values of $\Omega_m = 0.3$, $w = -1$.}
\end{deluxetable}

We generate simple toy data to validate the basic premise of the model. The data generation algorithm is described below:

\begin{enumerate}[1.]
	\item Draw a redshift from a power law distribution. For the low redshift survey this is $\mathcal{U}(0.0004, 0.01)^{0.5}$, and for the DES-like survey this is $\mathcal{U}(0.008, 1.0)^{0.33}$. \added{ For the low redshift survey, this is equivalent to sampling $y = z$ from $0.02$ to $\sqrt{0.1}$, for the high redshift survey, this is equivalent to sampling $y = z^2$ from $0.2$ to $1.0$. These distributions are arbitrary, and this test has been performed with various flat and power law distributions.}
	\item Draw a random mass probability from $\mathcal{U}(0, 1)$ and calculate the mass-brightness correction using $\delta(0) = 0.08$, $\delta(0)/\delta(\infty) = 0.5$, and equation \eqref{eq:mass}.
	\item Draw an absolute magnitude, stretch and color from the respective distributions $\mathcal{N}(-19.3, 0.1)$, $\mathcal{N}(0, 1)$, $\mathcal{N}(0, 0.1)$.
	\item Calculate $\mu(z)$ given the drawn redshift and cosmological parameters $\Omega_m = 0.3$, $w = -1$ under Flat $\Lambda$CDM cosmology. Use this to determine the true apparent magnitude of the object $m_B$ using $m_B = \mu + M_B - \alpha x_1 + \beta c$.
	\item Determine if the SN~Ia is detected using detection probability $P(S|m_B) = \mathcal{N}^{\rm skew}(13.72, 1.35, 5.87)$ for the low redshift survey (numeric values obtained by fitting to existing low redshift data). For the DES-like survey, accept with probability $P(S|m_B) = \Phi^C(23.14, 0.5)$. Repeat from step one until we have a supernova that passes. We use realistic values for the selection probability to ensure our model is numerically stable with highly skewed selection functions.
	\item Add independent, Gaussian observational error onto the true $m_B$, $x_1$, $c$ using Gaussian widths of $0.04$, $0.2$, $0.03$ respectively (following the mean uncertainty for DES-like SNANA simulations). Add extra color uncertainty in quadrature of $\kappa_0 + \kappa_1 z$, where $\kappa_0 = \kappa_1 = 0.03$.
\end{enumerate}

The selection functions parameters (a skew normal for low-redshift and a \replaced{complimentary}{complementary} error function for high-redshift) are all given independent uncertainty of $0.01$ (mean and width for the CDF selection function, and mean, width and skewness for the skew normal selection function). Draw from each survey simulation until we have 1000 low-$z$ supernovae and 1000 DES-like supernovae, representing a statistical sample of greater power than the estimated 350 supernovae for the DES-SN3YR sample. Sample data for 1000 high and low redshift supernovae are shown in Figure \ref{fig:simple_pop}, confirming the presence of strong selection effects in both toy surveys, as designed. 

We test four models: Flat $\Lambda$CDM, Flat $w$CDM, $\Lambda$CDM, and Flat $w$CDM with a prior $\Omega_m \sim \mathcal{N}(0.3, 0.01)$, with the latter included to allow sensitive tests on bias for $w$. To achieve statistical precision, we fit 100 realisations of supernovae datasets. Cosmological parameters are recovered without significant bias. Combined posterior surfaces of all 100 realisations fits for $\Lambda$CDM are shown in Figure \ref{fig:simple_ol} and fits for Flat $w$CDM are shown in Figure \ref{fig:simple_w}. By utilising the Stan framework and several efficient parametrisations (discussed further in Appendix \ref{app:optimisations}), fits to these simulations of 2000 supernovae take only on order of a single CPU-hour to run.

To investigate biases in the model in fine detail, we look for systematic bias in $\Omega_m$ in the Flat $\Lambda$CDM cosmology test, and bias in $w$ for the Flat $w$CDM test with strong prior $\Omega_m \sim \mathcal{N}(0.3, 0.01)$. This allows us to investigate biases without the investigative hindrances of non-Gaussian or truncated posterior surfaces. The strong prior on $\Omega_m$ cuts a slice through the traditional `banana' posterior surface in the $w$-$\Omega_m$ plane of Figure \ref{fig:simple_w}. Without making such a slice, the variation in $w$ is larger due to a shift along the degeneracy direction of the `banana'. By focusing the slice at an almost fixed $\Omega_m$, we can see the variation in the mean value of $w$ approximately perpendicular to the lines of degeneracy, instead of along them. The results of the analysis are detailed in Table \ref{tab:simple_model}, and demonstrate the performance of our model in recovering the true cosmological parameters.  As we are using 100 independent realisations, the precision of our determination of the mean simulation result is \added{ approximately a tenth of the quoted scatter (as a degree of non-Gaussianity of our fits will make this relationship inexact)}. The deviation from truth values is below this threshold, no significant bias is detected in either the Flat $\Lambda$CDM model or the Flat $w$CDM model. With this simple data, we also correctly recover underlying supernova populations, which can be seen in Figure \ref{fig:simple_w_super}.

\subsection{DES SN data validation}
\label{sec:simdes}

Many BHM methods have previously been validated on data constructed explicitly to validate the assumptions of the model. This is a useful consistency check that the model implementation is correct, efficient and free of obvious pathologies. However, the real test of a model is its application to realistic datasets that mimic expected observational data in as many possible ways. To this end, we test using simulations (using the SNANA package) that follow the observational schedule and observing conditions for the DES and low-$z$ surveys, where the low-$z$ sample is based on observations from CfA3 \citep{Hicken2009, Hicken2009a}, CfA4 \citep{Hicken2012} and CSP \citep{Contreras2010,Folatelli2010,Stritzinger2011}. Simulation specifics can be found in \citet{Kessler18}. \added{ The primary difference between the toy data of the previous section are the different underlying colour and stretch, inclusion of spectroscopic data cuts and light curve cuts, and inclusion of intrinsic dispersion models.}

Prior analyses often treated intrinsic dispersion simply as scatter in the underlying absolute magnitude of the underlying population \citep{Conley2011, Betoule2014}, but recent analyses require a more sophisticated approach. In our development of this model and tests of intrinsic dispersion, we analyse the effects of two different scatter models via simulations, the {\gten} and {\celeven} models described in Section \ref{sec:challenges}. The {\gten} models dispersion with 70\% contribution from coherent variation and 30\% from chromatic variation whilst the {\celeven} model has 25\% coherent scatter and 75\% from chromatic variation. These two broadband scatter models are converted to spectral energy distribution models for use in simulations in \citet{Kessler13}.

\begin{figure}
	\begin{center}
		\includegraphics[width=\columnwidth]{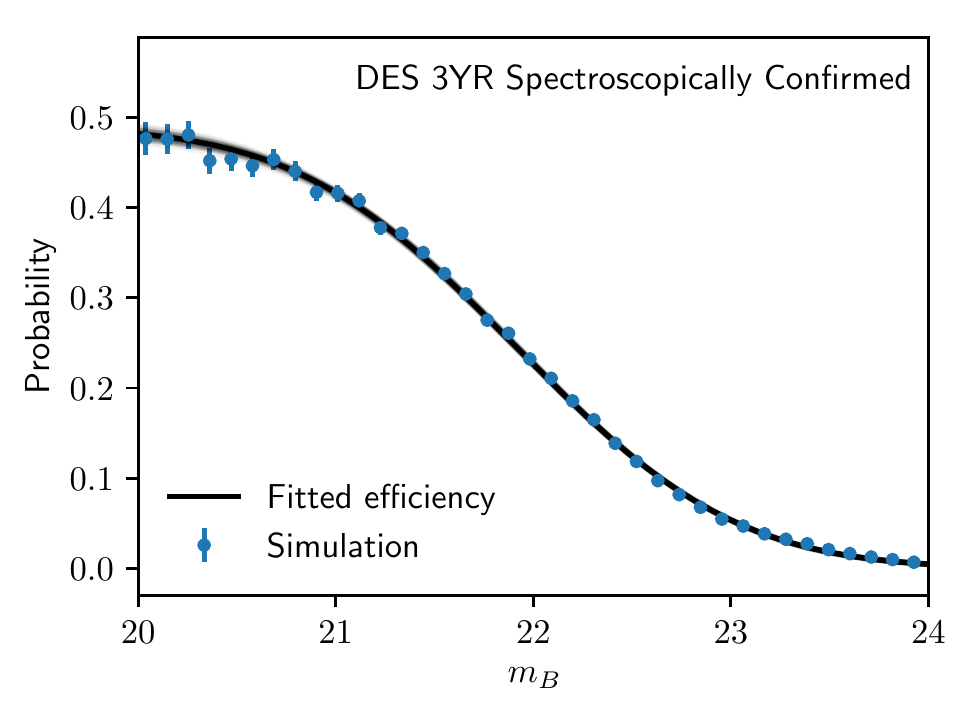}
		\includegraphics[width=\columnwidth]{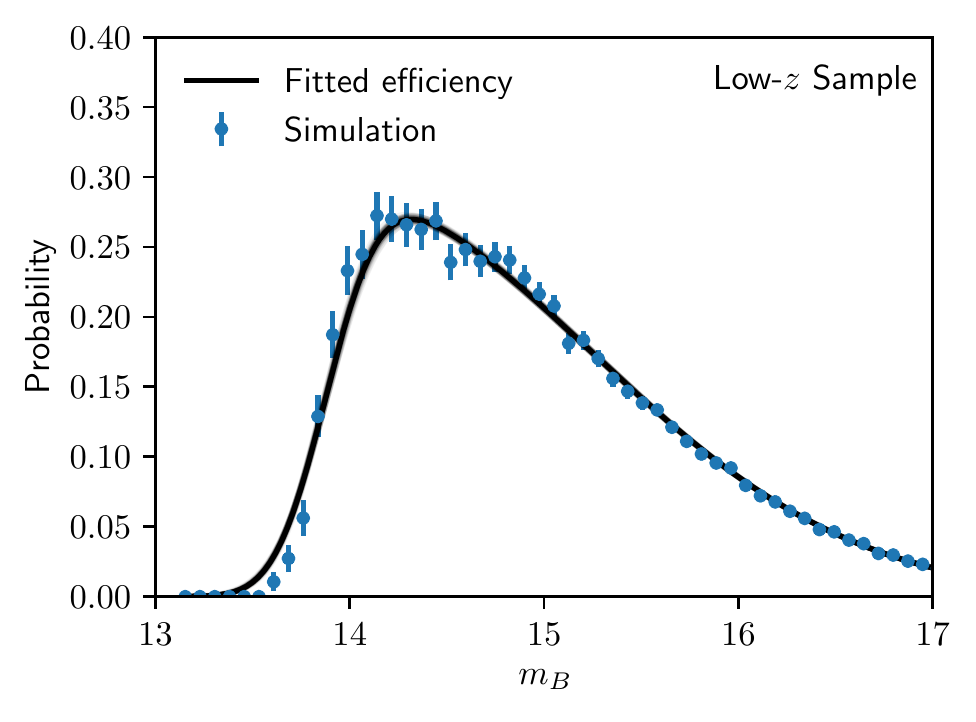}
	\end{center}
	\caption{Fitting the selection function for both the DES 3YR spectroscopically confirmed supernova sample and the low-$z$ sample. Blue errorbars represent the efficiency calculated by determining the ratio of discovered to generated supernovae in apparent magnitude bins for SNANA simulations. The black line represents the best fit analytic function for each sample, and the light grey lines surrounding the best fit value represent random realisations of analytic function taking into account uncertainty on the best fit value.}
	\label{fig:sf_snana}
\end{figure}

In addition to the improvements in testing multiple scatter models, we also include peculiar velocities for the low-$z$ sample, and our full treatment of systematics as detailed in \citet{Brout18SYS}. Our simulated populations are sourced from \citet[][hereafter {\sk}]{Scolnic2016} and shown in Table \ref{tab:dist}. Initial tests were also done with a second, Gaussian population with color and stretch populations centered on zero and with respective width $0.1$ and $1$, however cosmological parameters were not impacted by choice of the underlying population and we continue using only the {\sk} population for computational efficiency. The selection effects were quantified by comparing all the generated supernovae to those that pass our cuts, as shown in Figure~\ref{fig:sf_snana}. It is from this simulation that our analytic determination of the selection functions for the low-$z$ and DES survey are based. We run two simulations to determine the efficiency using the {\gten} and {\celeven} scatter models and find no difference in the functional form of the Malmquist bias between the two models. \added{ Uncertainty on the analytic selection function is incorporated into our fits, mitigating the imperfection of our analytic form by allowing it to vary in our fits.}

\begin{deluxetable}{lcccc}
	\renewcommand{\arraystretch}{1.2}
\tablewidth{\columnwidth}
	\label{tab:dist}
	\tablecomments{The {\sk} low-$z$ stretch distribution is formed as sum of two bifurcated Gaussians, with the mean and spread of each component given respectively.}
\tablecolumns{5}
\tablecaption{Supernova population distributions}
\tablehead {
	\colhead {Model} &
	\colhead { $\langle x_1 \rangle$  } &
	\colhead { $\sigma_{x_1}$ } &
	\colhead { $\langle c \rangle$ } &
	\colhead { $\sigma_c$ } 
}
	\startdata
		{\sk} low-$z$    &  $0.55$ \& $-1.5$ & $^{+0.45}_{-1.0}$ \& $^{+0.5}_{-0.5}$ & $-0.055$ & $^{+0.15}_{-0.023}$ \\
		{\sk} DES     & $0.973$ & $^{+0.222}_{1.472}$ & $-0.054$  &  $^{+0.101}_{0.043}$  \\
	\enddata
\end{deluxetable}

\begin{deluxetable*}{lrrrrr}
	\label{tab:standardisation}
	\tablewidth{\textwidth}
	\tablecolumns{6}
	\tablecaption{Realistic simulation standardisation parameters}
	\tablecomments{Standardisation parameters and base intrinsic scatter parameter results for the 100 fits to {\gten} and {\celeven} simulations. We show the average parameter mean and average standard deviation respectively, with the simulation scatter shown in brackets, such that each cell shows $\langle\mu\rangle\ [\langle\sigma\rangle\ ({\rm scatter})]$. The width of the intrinsic scatter ($\sigma_{m_B}$) does not have an input truth value as it is determined from the scatter model.}
	\tablehead{
		Model & $\alpha - \alpha_{\rm True}$ & $\beta - \beta_{\rm True}$ & $\langle M_B \rangle - \langle M_B \rangle_{\rm True}$ & $\sigma_{m_B}^{\rm DES}$ & $\sigma_{m_B}^{{\rm low-}z}$	
	}
	\startdata
		{\gten} Stat + Syst  & $0.022 \ [0.009\ (0.008)]$ & $ 0.34 \ [0.19\ (0.18)]$ & $-0.002 \ [0.028\ (0.015)]$ & $0.070 \ [0.022\ (0.018)]$ & $0.073 \ [0.025\ (0.022)]$\\
		{\gten}  Stat         & $0.000 \ [0.008\ (0.008)]$ & $0.33 \ [0.20\ (0.17)]$ & $0.001 \ [0.016\ (0.013)]$ & $0.069 \ [0.023\ (0.019)]$ & $0.072 \ [0.026\ (0.023)]$\\
		{\celeven} Stat + Syst  &  $0.002 \ [0.009\ (0.007)]$ & $-0.04 \ [0.15\ (0.13)]$ & $0.014 \ [0.030\ (0.018)]$ & $0.024 \ [0.016\ (0.011)]$ & $0.029 \ [0.020\ (0.014)]$ \\
		{\celeven} Stat         &  $0.000 \ [0.008\ (0.007)]$ & $-0.05 \ [0.16\ (0.13)]$ & $0.006 \ [0.016\ (0.015)]$ & $0.025 \ [0.016\ (0.012)]$ & $0.027 \ [0.020\ (0.015)]$ \\
	\enddata
\end{deluxetable*}

\begin{figure}
	\begin{center}
		\includegraphics[width=\columnwidth]{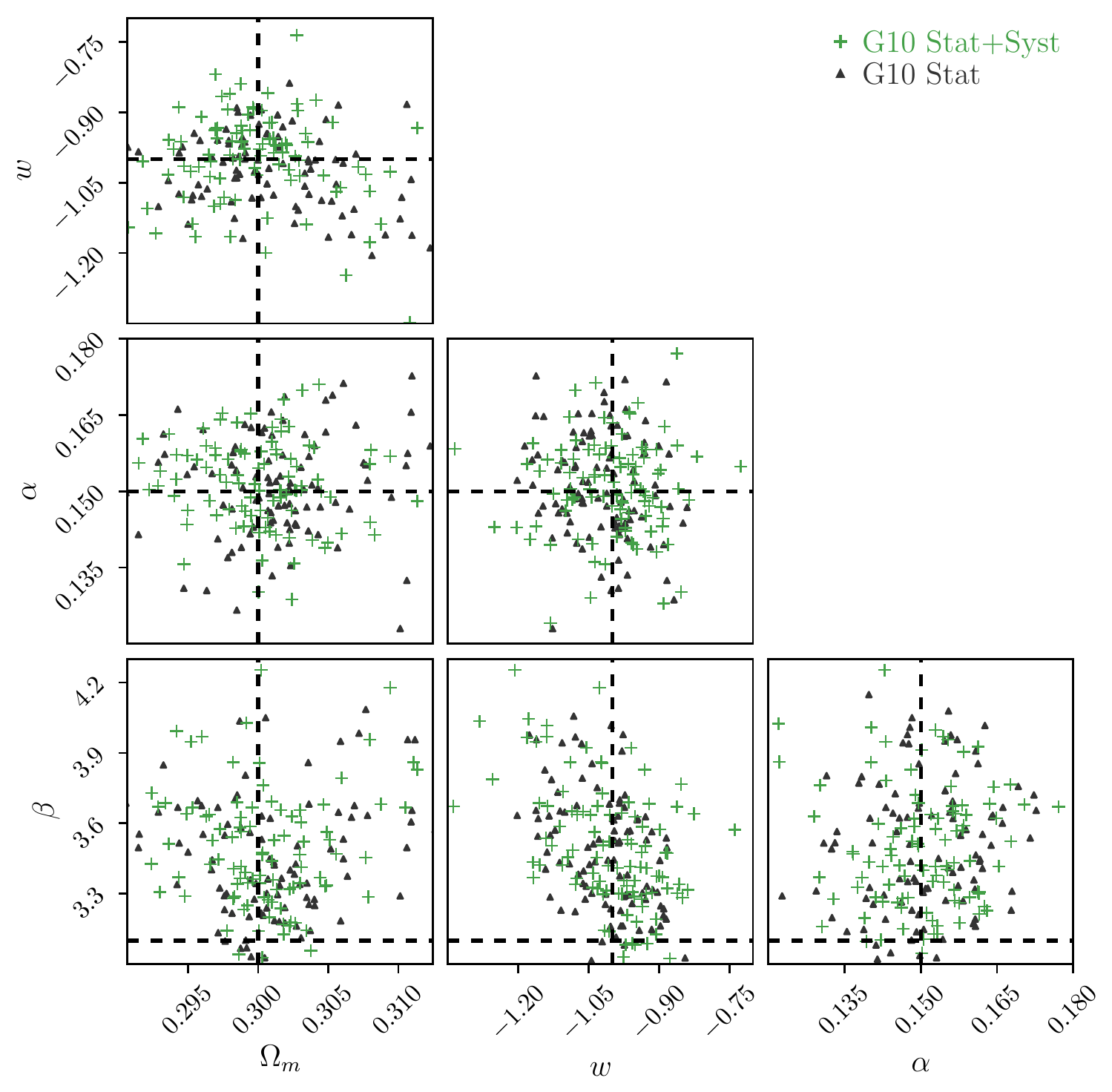}		\includegraphics[width=\columnwidth]{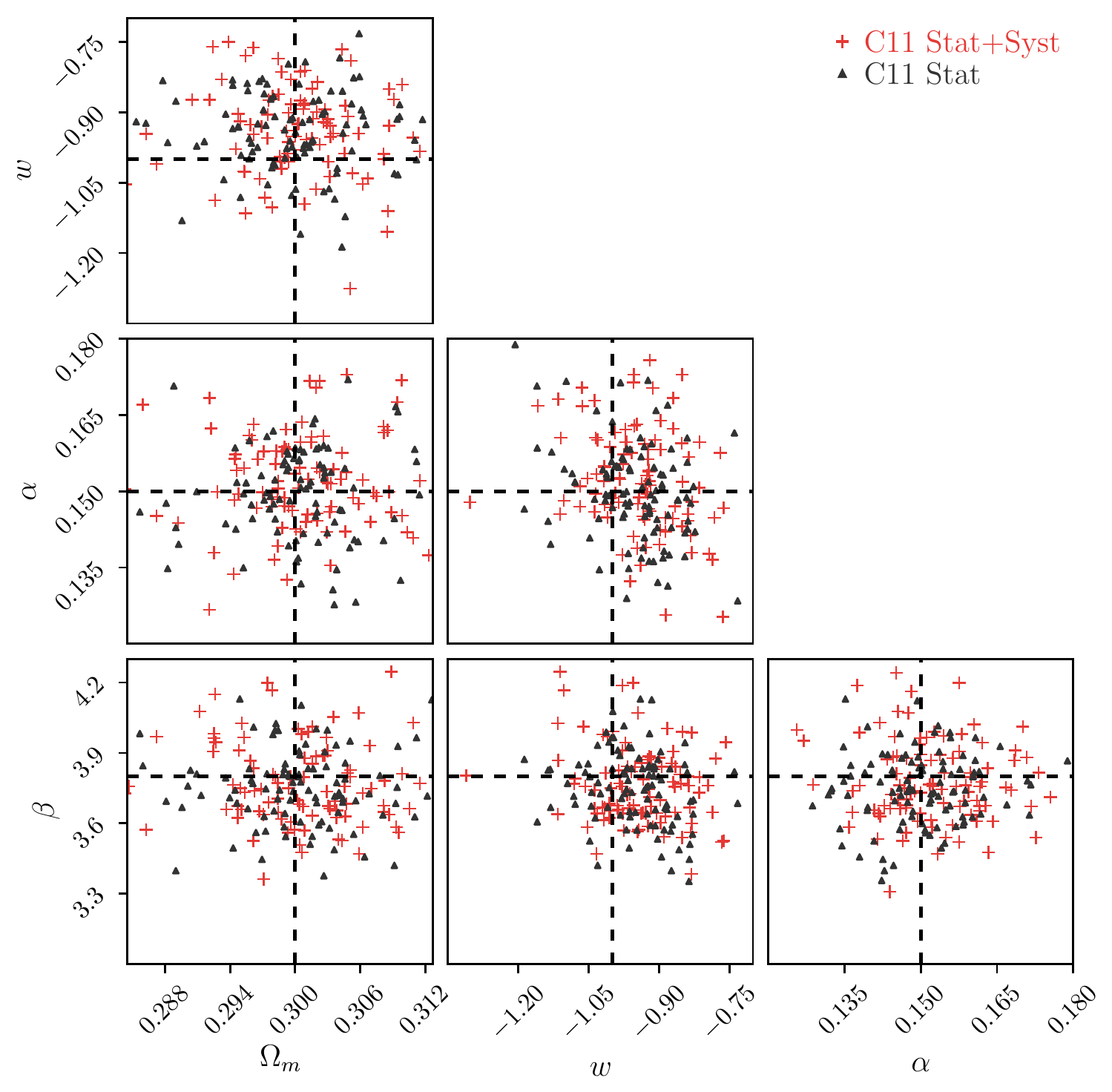}
	\end{center}
	\caption{Maximum posterior points for 100 realisations of supernova data for two intrinsic dispersion models - the {\gten} model for the top panel and the {\celeven} model for the bottom panel. Points are shown for parameters $\Omega_m$, $w$, $\alpha$ and $\beta$, with the other fit parameters being marginalised over. As we are unable to fully correct observed summary statistics, a step required by the lack of intrinsic scatter in the SALT2 model, we expect to see an offset in $\alpha$ and $\beta$. This in turn effects cosmology, resulting in small biases in $w$.}
	\label{fig:bulk_posterior}
\end{figure}

\begin{deluxetable}{lrr}
	\label{tab:bulk_summary}
	\tablecaption{Realistic simulation determination of $w$}
	\tablewidth{\columnwidth}
	\tablecomments{Investigating the combined 100 fits to {\gten} and {\celeven} simulations, fitting with both statistics only and also when including systematics. The quoted value for $w$ represents the average mean of the fits, with the average uncertainty being shown in square brackets and the simulation scatter (the standard deviation of the mean of 100 fits) shown in standard brackets. The bias significance represents our confidence that the deviation in the mean $w$ away from $-1$ is not due to statistical fluctuation.}
	\tablehead{		Model & $w\ \langle\mu\rangle\ [\langle\sigma_w\rangle$ (scatter)$]$ & $w$-Bias		}
	\startdata
		{\gten} Stat + Syst  &  $-0.998 \ [0.097\ (0.073)]$ & $(0.02\pm0.07)\sigma$  \\
		{\gten}  Stat         &  $-1.008\ [0.080\ (0.068)]$ & $(-0.10\pm0.08)\sigma$   \\
		{\celeven} Stat + Syst  &  $-0.945\ [0.098\ (0.077)]$ & $(0.55\pm0.08)\sigma$  \\
		{\celeven} Stat         &  $-0.948\ [0.079\ (0.066)]$ & $(0.65\pm0.08)\sigma$   \\
	\enddata
\end{deluxetable}

\begin{deluxetable}{crr}
	\label{tab:parameter_correlations}
	\renewcommand{\arraystretch}{0.6}
	\tablecaption{Reduced parameter correlations with $w$.}
	\tablecomments{Correlations determined from the combined 100 simulation fits. Correlations for the low-$z$ band systematics and the latent parameters representing selection function uncertainty are not shown but have negligible correlation. Zero superscripts indicate the Dark Energy Survey, and a superscript one represents the low-$z$ survey.}
	\tablehead{	\\	Parameter & {\gten} Stat+Syst & {\celeven} Stat+Syst }
	\startdata
		$\Omega_m$ &   $-$0.19 &     $-$0.21     \\
		$\alpha$ &   $-$0.17 &     $-$0.20     \\
		$\beta$ &   $-$0.29 &     $-$0.23     \\
		$\langle M_B \rangle$ &    0.68 &      0.66     \\
		$\delta(0)$ &    0.00 &      0.00     \\
		$\delta(\infty)/\delta(0)$ &    0.00 &      0.00     \\
		\\
		$\sigma_{\rm m_B}^{0}$ &    0.04 &      0.07     \\
		$\sigma_{\rm m_B}^{1}$ &    0.23 &      0.18     \\
		$\sigma_{x_1}^{0}$ &    0.04 &      0.03     \\
		$\sigma_{x_1}^{1}$ &    0.05 &      0.01     \\
		$\sigma_{c}^{0}$ &    0.01 &      0.11     \\
		$\sigma_{c}^{1}$ &    0.08 &      0.04     \\
		$\alpha_c^{0}$ &   $-$0.04 &      0.04     \\
		$\alpha_c^{1}$ &    0.03 &      0.01     \\
		$\kappa_{c0}^{0}$ &   $-$0.10 &     $-$0.05     \\
		$\kappa_{c0}^{1}$ &   $-$0.20 &     $-$0.17     \\
		$\kappa_{c1}^{0}$ &   $-$0.05 &     $-$0.01     \\
		$\kappa_{c1}^{1}$ &   $-$0.01 &      0.01     \\
		$\langle x_1^{0} \rangle$ &   $-$0.01 &     $-$0.05     \\
		$\langle x_1^{1} \rangle$ &   $-$0.02 &      0.02     \\
		$\langle x_1^{2} \rangle$ &   $-$0.04 &     $-$0.04     \\
		$\langle x_1^{3} \rangle$ &   $-$0.03 &     $-$0.06     \\
		$\langle x_1^{4} \rangle$ &   $-$0.06 &     $-$0.06     \\
		$\langle x_1^{5} \rangle$ &    0.04 &      0.02     \\
		$\langle x_1^{6} \rangle$ &    0.04 &      0.04     \\
		$\langle x_1^{7} \rangle$ &    0.08 &      0.03     \\
		$\langle c^{0} \rangle$ &   $-$0.05 &     $-$0.12     \\
		$\langle c^{1} \rangle$ &    0.11 &      0.03     \\
		$\langle c^{2} \rangle$ &    0.11 &      0.06     \\
		$\langle c^{3} \rangle$ &    0.14 &      0.04     \\
		$\langle c^{4} \rangle$ &   $-$0.11 &     $-$0.11     \\
		$\langle c^{5} \rangle$ &   $-$0.15 &     $-$0.08     \\
		$\langle c^{6} \rangle$ &   $-$0.12 &     $-$0.13     \\
		$\langle c^{7} \rangle$ &   $-$0.12 &     $-$0.06     \\
		\\
		$\delta [ {\rm SALT}_0 ]$ &    0.05 &      0.05     \\
		$\delta [ {\rm SALT}_1 ]$ &   $-$0.01 &      0.02     \\
		$\delta [ {\rm SALT}_2 ]$ &   $-$0.10 &     $-$0.09     \\
		$\delta [ {\rm SALT}_3 ]$ &   $-$0.03 &     $-$0.03     \\
		$\delta [ {\rm SALT}_4 ]$ &    0.08 &      0.09     \\
		$\delta [ {\rm SALT}_5 ]$ &    0.01 &      0.02     \\
		$\delta [ {\rm SALT}_6 ]$ &    0.05 &      0.07     \\
		$\delta [ {\rm SALT}_7 ]$ &   $-$0.11 &     $-$0.10     \\
		$\delta [ {\rm SALT}_8 ]$ &    0.01 &      0.02     \\
		$\delta [ {\rm SALT}_9 ]$ &    0.02 &      0.02     \\
		$\delta [ {\rm MWE}_{B-V} ]$ &    0.03 &      0.02     \\
		$\delta [ {\rm HST Calib} ]$ &   $-$0.07 &     $-$0.07     \\
		$\delta [ v_{\rm pec} ]$ &    0.00 &     $-$0.01     \\
		$\delta [ {\rm \delta z } ]$ &    0.01 &      0.00     \\
		$\delta [ \Delta g ]$ &    0.05 &      0.11     \\
		$\delta [ \Delta r ]$ &    0.16 &      0.10     \\
		$\delta [ \Delta i ]$ &   $-$0.16 &     $-$0.18     \\
		$\delta [ \Delta z ]$ &   $-$0.26 &     $-$0.26     \\
		$\delta [ \Delta \lambda_g ]$ &    0.16 &      0.20     \\
		$\delta [ \Delta \lambda_r ]$ &    0.05 &      0.06     \\
		$\delta [ \Delta \lambda_i ]$ &    0.00 &     $-$0.01     \\
		$\delta [ \Delta \lambda_z ]$ &    0.09 &      0.07     \\
	\enddata
\end{deluxetable}

\begin{figure*}
	\begin{center}
		\includegraphics[width=\textwidth]{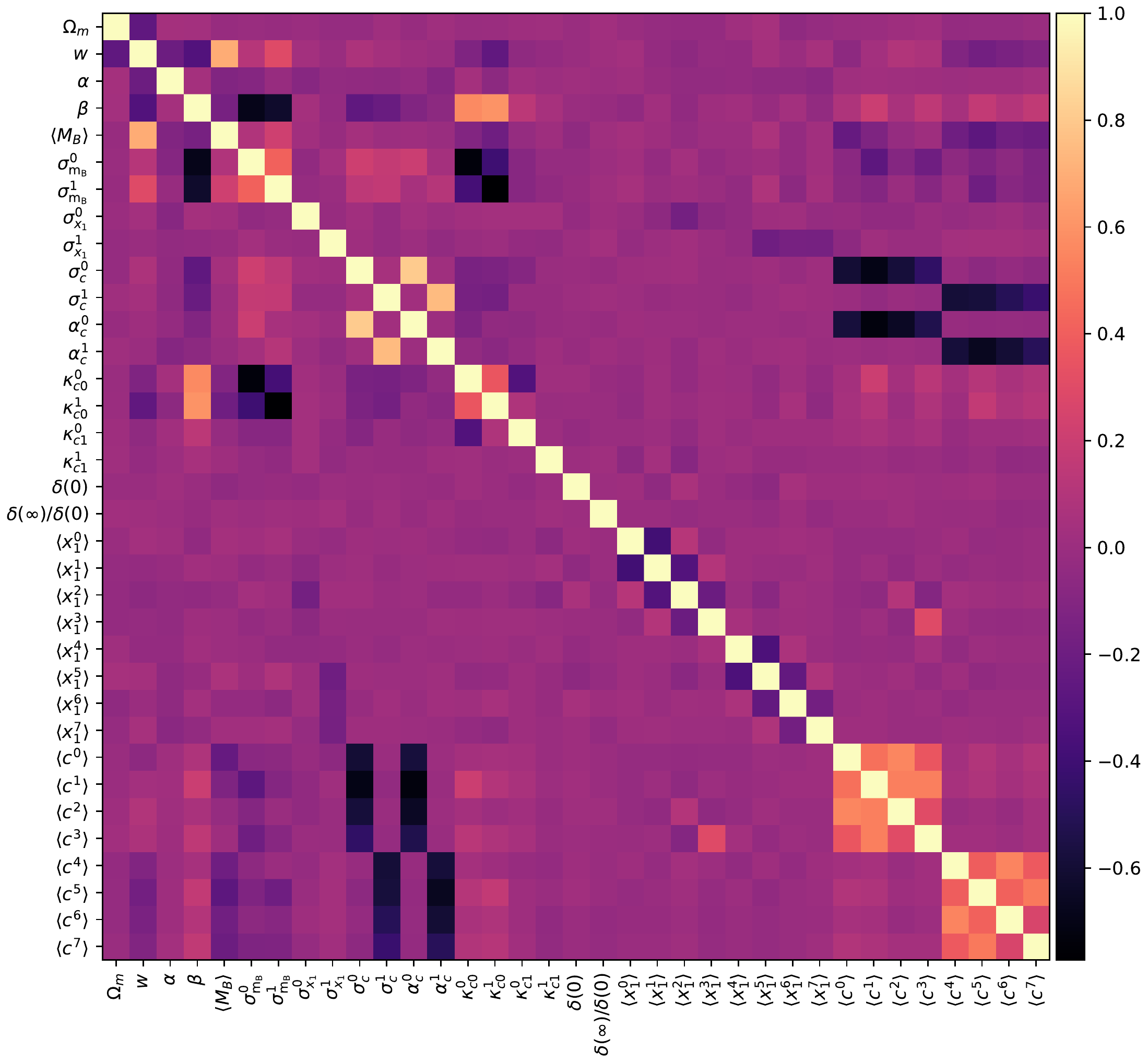}
	\end{center}
	\caption{Parameter correlations for the combined fits to the 100 {\gten} scatter model simulations. We see that the primary correlations with $w$ enter through $\alpha$, $\beta$ and $\langle M_B \rangle$, as shown in Table \ref{tab:parameter_correlations}. Whilst $\langle M_B \rangle$ is generally thought to be a nuisance parameter, we find cosmological correlation. We note that, by fixing $H_0$ in our distance modulus calculation, $\langle M_B \rangle$ absorbs any cosmological uncertainty on this term. Additionally $\langle M_B \rangle$ also effects the selection efficiency, which was computed from simulations with a fixed $M_B$ value, introducing a second plausible source of correlation. Also visible in this figure are several other interesting relationships. $\beta$ is strongly anti-correlated with intrinsic dispersion $\sigma_{m_B}$ for both surveys (DES-like and low-$z$), with $\sigma_{m_B}$ showing strong anti-correlation with $\kappa^0_c$. This relationship is indeed expected --- as $\kappa^0_c$ grows larger (more unexplained dispersion on the color observation), the width of the supernova population in apparent magnitude space increases. As the fit prefers it to conform to the observed width of the distribution, the extra width in color causes the inherent magnitude smearing amount to decrease. And with extra freedom on the observed color from $\kappa^0_c$, $\beta$ shifts in response. The other striking feature in the plot is the strong correlation blocks in the bottom right and the anti-correlation stripes on the edges. These too are expected, for they show the relationship between the color distribution's mean value, its width and its skewness. As skewness or population width increases, the effective mean of the population shifts (see Appendix \ref{app:approx} for details), creating anti-correlation between skewness and the (Gaussian) mean color population. Strong anti-correlation between $\kappa_{c0}^0$ and $\kappa_{c0}^1$ with $\sigma_{m_B}$ reveals the strong population degeneracy, and -- for the C11 simulation results -- a constrained positive value shows that a finite non-zero extra color dispersion is indeed preferred by our model. }
	\label{fig:bulk_cor}
\end{figure*}

Each realisation of simulated SN~Ia light curves contains the SALT2 light-curve fits and redshifts to 128 low-$z$ supernovae, and 204 DES-like supernovae, such that the uncertainties found when combining chains is representative of the uncertainty in the DES-SN3YR sample. As our primary focus is Dark Energy, we now focus specifically on the Flat $w$CDM model with matter prior.

Points of maximum posterior for 100 data realisations are shown in Figure \ref{fig:bulk_posterior}. The parameter bounds and biases for $w$ are listed in Table \ref{tab:bulk_summary},  and secondary parameters are shown in Table \ref{tab:standardisation}.

Table \ref{tab:bulk_summary} shows that the {\gten} model is consistent with $w=-1$, whilst the {\celeven} model show evidence of bias on $w$, scattering high. However, their deviation from the truth value represents a shift of approximately $0.5\sigma$ when taking into account the uncertainty on fits to $w$. The bias is sub-dominant to both the size of the uncertainty for each fit, and the scatter induced by statistical variance in the simulations. We also note that the simulations do not vary cosmological parameters nor population. As our model does include uncertainty on those values, the simulation scatter is expected to be less than the model uncertainty, and represents a minimum bound on permissible uncertainty values. 

Table \ref{tab:standardisation} shows a clear difference in both $\beta$ and $\sigma_{m_B}$ across the {\gten} and {\celeven} simulations. As expected, the {\celeven} simulations recover a far smaller intrinsic magnitude scatter, giving results of approximately $0.025$ when compared to the result of $0.070$ for the {\gten} simulations. The extra smearing of the {\celeven}  model does not result in a significantly biased $\beta$ value compared to the average uncertainty on $\beta$, with recovery of $\beta\approx3.76$ close to the input truth value of $3.8$, however the $\beta$ recovery for the {\gten} simulations is biased high, finding $\beta\approx3.44$ with an input of $3.1$. Interestingly, $w$-bias is only found for the {\celeven} simulations. A measure of the significance of the parameter bias can be calculated by comparing the bias to a tenth of the scatter (as our Monte-Carlo estimate uncertainty is $\sqrt{100}$ of the scatter). From this, we can see that most biases are detected with high statistical significance due to the large number of simulations tested against. 
 
We investigate the cosmological bias and find its source to be a bias in the observed summary statistics (i.e. the $\hat{m}_B$, $\hat{x}_1$, $\hat{c}$ output from SALT2 light curve fitting), in addition to incorrect reported uncertainty on the summary statistics. To confirm this, we run two tests. The first of which, we replace the SALT2-fitted $\hat{m}_B$, $\hat{x}_1$ and $\hat{c}$ with random numbers drawn from a Gaussian centered on the true SALT2 $m_B$, $x_1$ and $c$ values with covariance as reported by initial light curve fits. With this test, both the {\gten} and {\celeven} fits recover $w=-1.00$ exactly. Our second test aims to test our model whilst allowing biases in the summary statistics not caused from intrinsic scatter through. \added{ That is, the first test ascertained that biases in the summary statistics are the cause of cosmological bias. It is thus important to determine the source of those biases; whether they are from intrinsic scatter model or another aspect of the simulation. To} this end, we test a set of 100 simulations generated using an intrinsic dispersion model of only coherent magnitude scatter. \added{ We find $w=-1.00$, showing} that the source of the biases in summary statistics is the underlying intrinsic scatter model. From this, the main challenge of improving our methodology is to handle the fact that observational uncertainty reported from fitting the SALT2 model to light curves is incorrect, non-Gaussian and biased. Our current model and techniques can quantify the effect of different scatter models on biasing the observed summary statistics, but being unable to constrain the `correct' (simulated) scatter model in our model fit means we cannot fully correct for the bias introduced by an unknown scatter model. 

Unfortunately, adding extra fit parameters to allow for shifting observables washes out our ability to constrain cosmology, and applying a specific bias correction requires running a fiducial simulation (assuming cosmology, population and scatter model), which presents difficulties when trying to account for correlations with population and scatter model. This is compounded by the fact that bias corrections do not in general improve fits (increase the log posterior), and so are difficult to fit parametrically. Works such as \citet{Kessler2017} show that bias corrections can be applied to supernovae datasets that can robustly handle multiple intrinsic scatter models, and future work will center on uniting these methodologies --- incorporating better bias corrections that separate intrinsic scatter bias and non-Gaussian summary statistic bias from Malmquist bias, without having to precompute standardisation parameters and populations.

Difficulty in providing an adequate parametrisation for realistic intrinsic dispersion, and the simplification of Malmquist bias to only apparent magnitude also leads to biases in the population parameters. \added{ To determine if the underlying population mischaracterisation was a cause of cosmological bias, we ran fits wherein the underlying population was fixed to the known distributions used for the simulation. These fits did not change the bias in the {\celeven} simulation. We conclude that the biased population recovery is not the cause of cosmological bias.} As the population parameters recovered using the simplistic toy supernova data in the previous section do not exhibit significant bias, future work will focus on intrinsic dispersion and Malmquist bias rather than alternate parametrisations of the underlying supernova population.

Table \ref{tab:parameter_correlations} lists the fit correlations between our model fit parameters (excluding the low-$z$ band systematics, and Malmquist bias uncertainty parameters which had negligible correlation), showing (in order) cosmological parameters, standardisation parameters, population width and skewness parameters, intrinsic dispersion parameters, mass-step parameters, population mean parameters, SALT2 model systematics, dust systematic, global HST calibration systematic, peculiar velocity systematic, global redshift systematic and DES band magnitude and wavelength systematics. Figure \ref{fig:bulk_cor} show the full correlations between all non-systematic model parameters. Other interesting correlations are shown and discussed in Figure \ref{fig:bulk_cor}. The band systematics for DES filters $g$, $r$ and $i$ also show significant correlation with $w$, highlighting the importance of minimising instrumental uncertainty.

For the sample size of the DES $+$ low-$z$ supernova samples (332 supernova), the bias from intrinsic scatter models is sub-dominant to the statistical uncertainty, as shown in Figure \ref{fig:bulk_posterior}. For our full systematics model, the bias represents a deviation between $0\sigma$ to $0.5\sigma$ depending on scatter model, and given that they remain sub-dominant, we will leave more complicated treatment of them for future work.

\subsection{Uncertainty Analysis}
\label{sec:ucnert}

\begin{deluxetable*}{llrr}
	\label{tab:uncert}
	\tablecaption{$w$ error budget}
	\tablecomments{Error budget determined from analysing uncertainty on simulation data whilst progressively enabling model features. We start from the top of the table, only varying cosmological parameters $\Omega_m$ and $w$, and then progressively unlock parameters and let them fit as we progress down the table. The cumulative uncertainty shows the total uncertainty on $w$ on the fit for all, where the $\sigma_w$ term is derived by taking the quadrature difference in cumulative uncertainty as we progress.}
	\tablehead{		Feature &  Parameters & $\sigma_w$ & Cumulative }
	\startdata
		Cosmology                      & $\Omega_m$, $w$ & 0.051  &   0.051       \\
		Standardisation                & $\alpha$, $\beta$, $\langle M_B \rangle$, $\delta(0)$, $\delta(\infty)/\delta(0)$ & 0.046  &   0.068       \\
		Intrinsic scatter              & $\kappa_0$, $\kappa_1$ & 0.020  &   0.071       \\
		Redshift-independent populations             & $\sigma_{M_B}$, $\sigma_c$, $\sigma_{x_1}$, $\alpha_c$ & 0.022  &   0.074       \\
		Redshift-dependent populations & $\langle c_i \rangle$, $\langle x_{1,i} \rangle$ & 0.030  &   0.080   \\
		Systematics                    & $\delta \mathcal{Z}_i$, $\delta S$ & 0.054  &   0.096      \\ 
	\enddata
\end{deluxetable*}

With the increased flexibility of Bayesian hierarchical models over traditional models, we expect to find an increased uncertainty on parameter inference. \added{This increased uncertainty is one of the strengths of hierarchical models as it represents a more thorough accounting of model uncertainty.} To characterise the influence of the extra degrees of freedom in our model, we analyse the uncertainty on $w$ averaged across 10 nominal simulations of the DES-SN3YR sample with various model parameters allowed to either vary or stay locked to a fixed value. By taking the difference in uncertainty in quadrature, we can infer the relative contribution for each model feature to the uncertainty error budget.

The error budget detailed in Table \ref{tab:uncert} shows that our uncertainty is dominated by statistical error, as the total statistical uncertainty is on $w$ is $\pm0.08$. With the low number of supernovae in the DES-SN3YR sample, this is expected. We note that the label `Systematics' in Table \ref{tab:uncert} represents all numerically computed systematics (as discussed in Section \ref{sec:systreat}) and systematic uncertainty on the selection function.

\subsection{Methodology Comparison}
\label{sec:comp}

We compare the results of our model against those of the BBC+CosmoMC method \citep{Kessler2017}. BBC+CosmoMC has been used in prior analyses, such as the Pantheon sample analysis of \citet{Scolnic2017} and is being used in the primary analysis of the DES-SN3YR sample \citep{DESKEY}. The BBC method is a two-part process, BBC computes bias corrections for observables, and then the corrected distances are fit using CosmoMC \citep{lewis2002cosmomc}. For shorthand, we refer to this combined process as the BBC method hereafter in this paper, as we are concerned with the results of cosmological parameter inference. As a leading supernova cosmology method, it provides a good consistency check as to the current levels of accuracy in recovering cosmological parameters.

\begin{figure}
	\begin{center}
		\includegraphics[width=\columnwidth]{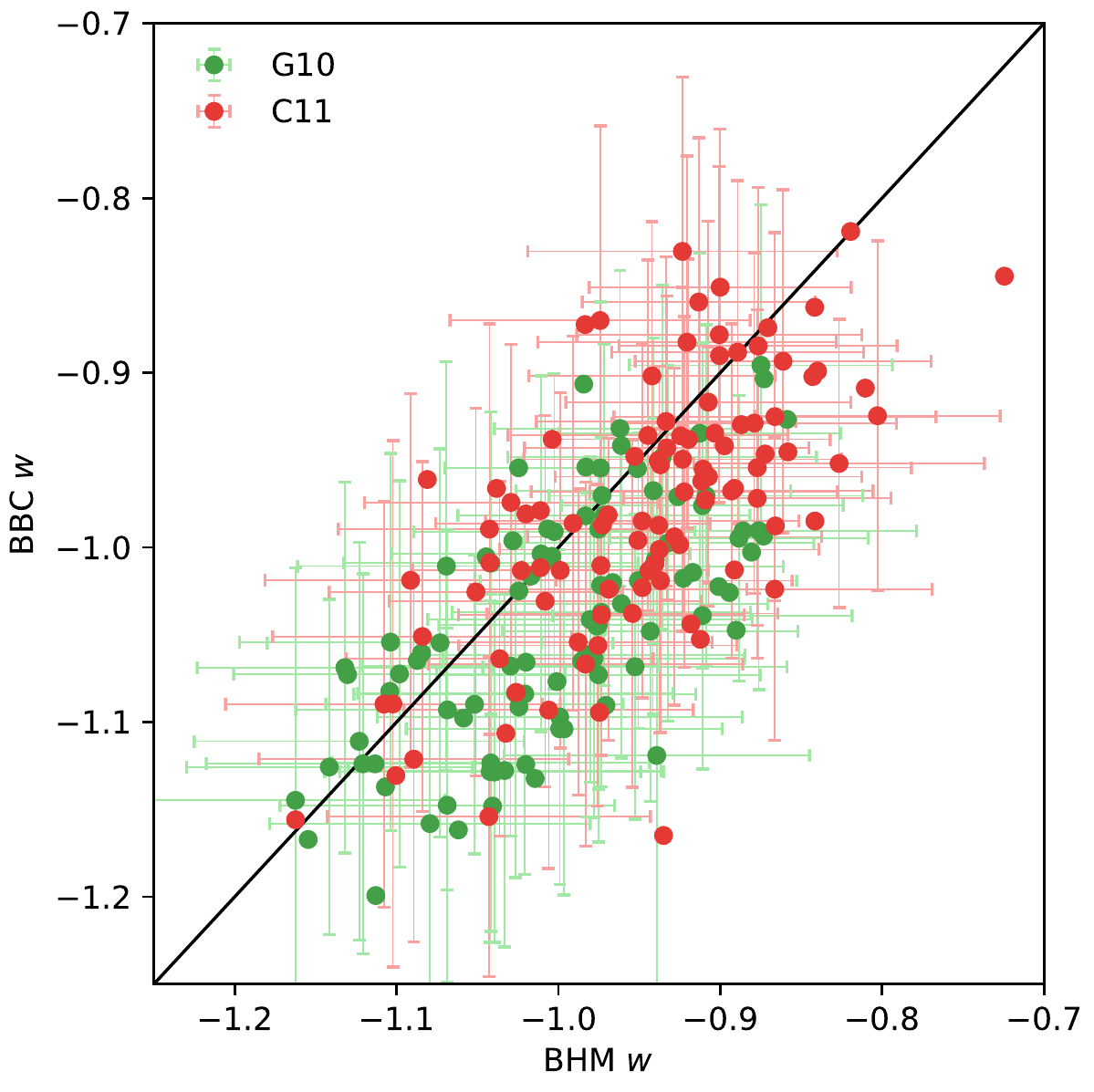}
	\end{center}
	\caption{Recovered $w$ for the 200 validation simulations with full treatment of statistical and systematic errors. Uncertainty on the recovered $w$ value is shown for every second data point for visual clarity.}
	\label{fig:bulk_comparison}
\end{figure}

To this end, we take the results of the BBC method which were also run on the same set of 200 validation simulations and compare the recovered $w$ values to those of our method. The results are detailed in Table \ref{tab:bulk_summary_bbc}, and a scatter plot of the simulation results is presented in Figure \ref{fig:bulk_comparison}. 

As shown in \citet{Brout18SYS}, the BBC method recovers cosmological parameters without bias so long as the intrinsic scatter model is known. As we do not know the correct intrinsic scatter model, the BBC method averages the results when using bias corrections from {\gten} and from {\celeven}. As such, we expect the BBC method to have a $w$-bias in one direction for {\gten} simulations and the other direction for {\celeven} simulations. These results are consistent with those displayed in Table \ref{tab:bulk_summary_bbc}. Both BBC method and {\steve} are sensitive to the intrinsic scatter model, finding differences of $\sim0.066$ and $0.053$ respectively in $w$ when varying the scatter model. The BBC method finds $w$ biased low for G10 and $w$ biased high for C11 (by about $\pm0.03$), so taking the average result only results in a small bias of $-0.01$ in $w$. Our method shows a small improvement in the insensitivity to the intrinsic scatter model (having a decrease in difference in $w$ between the {\gten} and {\celeven} models), finding no bias for {\gten} but a $w$ biased high for {\celeven}. This decrease in error is not statistically significant as we have statistical uncertainty of $\sim0.01$ for $100$ simulation realisations. The average bias over the two scatter models is $+0.028$, representing a larger bias than the BBC method.

When comparing both the {\gten} and {\celeven} set of simulations independently, our model differs from BBC in its average prediction of $w$ by $+0.044$ and $+0.033$ respectively. For the {\gten} model this difference is a result of bias in the BBC results, however for the {\celeven} simulations this is a result of both bias from BBC, and a larger bias from our method. These results also allow us to state the expected values for $w$ when run on the DES-SN3YR sample. When using Planck priors our uncertainty on $w$ is reduced compared to using our simulation Gaussian prior on $\Omega_m$, shrinking the average $w$-difference from $0.06$ to $0.04$. After factoring this into our uncertainty, we expect our BHM method to, on average, recover $w_{\rm BHM} = w_{\rm BBC} +0.04\pm0.04$.

Having established that our method exhibits similar shifts in the recovery of $w$ compared to BBC, future work will focus on improving the parametrisation of intrinsic scatter model into our framework, with the goal of minimising the effect of the underlying scatter model on the recovery of cosmology.

\begin{deluxetable}{lccc}
	\tablecaption{$w$ bias comparison}
	\tablecomments{We characterise the bias on $w$ using the 100 simulations for the {\gten} scatter model and 100 simulations for {\celeven} scatter model. We also show the results when combining the {\gten} and {\celeven} models into a combined set of 200 simulations. The mean $w$ value for our method and BBC are presented, along with the mean when averaging the difference between our method and BBC for each individual simulation. Averages are computed giving each simulation sample the same weight. In the model, $\Delta$ represents Steve - BBC. The final row shows the scatter between Steve and BBC for the different simulations.}
	\tablehead{		Model & {\gten} & {\celeven} & ({\gten} + {\celeven}) }
	\startdata
		Steve $\langle w \rangle$           & $-0.998\pm0.007$ & $-0.945\pm0.007$ & $-0.972\pm0.006$\\
		BBC $\langle w \rangle$             & $-1.044\pm0.006$ & $-0.978\pm0.007$ & $-1.010\pm0.005$\tablenotemark{a}\\
		$\Delta$  $\langle w \rangle$  & $+0.044\pm0.006$ & $+0.033\pm0.006$ & $+0.038\pm0.004$ \\
		$\Delta$  $\sigma_w$           & $0.057\pm0.004$ & $0.062\pm0.004$ & $0.060\pm0.003$ \\
	\enddata
	\tablenotetext{a}{This value is computed with each simulation having the same weight. It disagrees with the value provided in \citet[Table 10, row 3]{Brout18SYS} which uses inverse variance weighted averages. We do not utilise this weight because the variance is correlated with the value of $w$ due to the $\Omega_m$ prior applied in the fitting process. We note that if inverse variance weighting is applied to both datasets, they both shift by $\Delta w \approx 0.005$, and thus the predicted difference between the BBC method and {\steve} remains the same.}
	\label{tab:bulk_summary_bbc}
\end{deluxetable}

\section{Conclusions}
\label{sec:conclusion}

In this paper we have outlined the creation of a hierarchical Bayesian model for supernova cosmology. The model takes into account selection effects and their uncertainty, fits underlying populations and standardisation parameters, incorporates unexplained dispersion from intrinsic scatter color smearing and incorporates uncertainty from peculiar velocities, survey calibration, HST calibration, dust, a potential global redshift offset, and SALT2 model uncertainty. Furthermore, our uncertainties in standardisation, population, mass-step and more, being explicitly parametrised in our model, are captured with covariance intact, an improvement on many previous methods. The model has been optimised to allow for hundreds of supernovae to be modelled fully with latent parameters. It runs in under an hour of CPU time and scales linearly with the number of supernovae, as opposed to polynomial complexity of matrix inversion of other methods.

The importance of validating models using high-precision statistics gained by performing fits to hundreds of data realisations cannot be overstated, however this validation is lacking in many earlier BHM models for supernova cosmology. We have validated this model against many realisations of simplistic simulations with well-known and well-defined statistics, and found no cosmological bias. When validating using SNANA simulations, we find evidence of cosmological bias which is traced back to light curve fits reporting biased observables and incorrect covariance. Allowing fully parametrised corrections on observed supernovae summary statistics introduces too many degrees of freedom and is found to make cosmology fits too weak. Allowing simulation based corrections to vary in strength is found to give minor reductions in $w$ bias, however the uncertainty on the intrinsic scatter model itself limits the efficacy of the bias corrections. For the data size represented in the DES three-year spectroscopic survey, the determined biases should be sub-dominant to other sources of uncertainty, however this cannot be expected for future analyses with larger datasets. Stricter bias corrections calculated from simulations are required to reduce bias. Ideally, this would include further work on the calculation of intrinsic dispersion of the type Ia supernova population such that we can better characterise this bias.

With our model being validated against hundreds of simulation realisations, representing a combined dataset over more than $60\,000$ simulated supernovae, we have been able to accurately determine biases in our model and trace their origin. With the current biases being sub-dominant to the total uncertainty, we now prepare to analyse the DES three-year dataset.

\section*{Acknowledgements}

Plots of posterior surfaces and parameter summaries were created with \verb|ChainConsumer| \citep{Hinton2016}.

Funding for the DES Projects has been provided by the U.S. Department of Energy, the U.S. National Science Foundation, the Ministry of Science and Education of Spain, 
the Science and Technology Facilities Council of the United Kingdom, the Higher Education Funding Council for England, the National Center for Supercomputing 
Applications at the University of Illinois at Urbana-Champaign, the Kavli Institute of Cosmological Physics at the University of Chicago, 
the Center for Cosmology and Astro-Particle Physics at the Ohio State University,
the Mitchell Institute for Fundamental Physics and Astronomy at Texas A\&M University, Financiadora de Estudos e Projetos, 
Funda{\c c}{\~a}o Carlos Chagas Filho de Amparo {\`a} Pesquisa do Estado do Rio de Janeiro, Conselho Nacional de Desenvolvimento Cient{\'i}fico e Tecnol{\'o}gico and 
the Minist{\'e}rio da Ci{\^e}ncia, Tecnologia e Inova{\c c}{\~a}o, the Deutsche Forschungsgemeinschaft and the Collaborating Institutions in the Dark Energy Survey. 

The Collaborating Institutions are Argonne National Laboratory, the University of California at Santa Cruz, the University of Cambridge, Centro de Investigaciones Energ{\'e}ticas, 
Medioambientales y Tecnol{\'o}gicas-Madrid, the University of Chicago, University College London, the DES-Brazil Consortium, the University of Edinburgh, 
the Eidgen{\"o}ssische Technische Hochschule (ETH) Z{\"u}rich, 
Fermi National Accelerator Laboratory, the University of Illinois at Urbana-Champaign, the Institut de Ci{\`e}ncies de l'Espai (IEEC/CSIC), 
the Institut de F{\'i}sica d'Altes Energies, Lawrence Berkeley National Laboratory, the Ludwig-Maximilians Universit{\"a}t M{\"u}nchen and the associated Excellence Cluster Universe, 
the University of Michigan, the National Optical Astronomy Observatory, the University of Nottingham, The Ohio State University, the University of Pennsylvania, the University of Portsmouth, 
SLAC National Accelerator Laboratory, Stanford University, the University of Sussex, Texas A\&M University, and the OzDES Membership Consortium.

Based in part on observations at Cerro Tololo Inter-American Observatory, National Optical Astronomy Observatory, which is operated by the Association of 
Universities for Research in Astronomy (AURA) under a cooperative agreement with the National Science Foundation.

The DES data management system is supported by the National Science Foundation under Grant Numbers AST-1138766 and AST-1536171.
The DES participants from Spanish institutions are partially supported by MINECO under grants AYA2015-71825, ESP2015-66861, FPA2015-68048, SEV-2016-0588, SEV-2016-0597, and MDM-2015-0509, 
some of which include ERDF funds from the European Union. IFAE is partially funded by the CERCA program of the Generalitat de Catalunya.
Research leading to these results has received funding from the European Research
Council under the European Union's Seventh Framework Program (FP7/2007-2013) including ERC grant agreements 240672, 291329, and 306478.
We  acknowledge support from the Australian Research Council Centre of Excellence for All-sky Astrophysics (CAASTRO), through project number CE110001020, and the Brazilian Instituto Nacional de Ci\^encia
e Tecnologia (INCT) e-Universe (CNPq grant 465376/2014-2).

This manuscript has been authored by Fermi Research Alliance, LLC under Contract No. DE-AC02-07CH11359 with the U.S. Department of Energy, Office of Science, Office of High Energy Physics. The United States Government retains and the publisher, by accepting the article for publication, acknowledges that the United States Government retains a non-exclusive, paid-up, irrevocable, world-wide license to publish or reproduce the published form of this manuscript, or allow others to do so, for United States Government purposes.



\bibliographystyle{mnras}
\bibliography{bib}


\appendix

\section{Selection Effect Derivation}
\label{app:selection}

\subsection{General Selection Effects}
\label{app:selection1}
When formulating and fitting a model using a constraining dataset, we wish to resolve the posterior surface defined by
\begin{equation}
P(\theta | {\rm data}) \propto P({\rm data} | \theta) P(\theta),
\end{equation}
which gives the probability of the model parameter values ($\theta$) given the data.  Prior knowledge of the allowed values of the model parameters is encapsulated in the prior probability $P(\theta)$. Of primary interest to us is the likelihood of observing the data given our parametrised model, $\mathcal{L} \equiv P({\rm data} | \theta)$. When dealing with experiments that have imperfect selection efficiency, our likelihood must take that efficiency into account.  We need to describe the probability that the events we observe are both drawn from the distribution predicted by the underlying theoretical model \textit{and} that those events, given they happened, are subsequently successfully observed.  To make this extra conditional explicit, we write the likelihood of the data given an underlying model, $\theta$, \textit{and} that the data are included in our sample, denoted by $S$, as:
\begin{equation}
\mathcal{L}(\theta; {\rm data}) = P({\rm data} | \theta, S). \label{eq:app_like}
\end{equation}
A variety of selection criteria are possible, and in our method we use our data in combination with the proposed model to determine the probability of particular selection criteria.  That is, we characterise a function $P(S|{\rm data},\theta)$, which colloquially can be stated as \textit{the probability of a potential observation passing selection cuts, given our observations and the underlying model}. We can introduce this expression in a few lines due to symmetry of joint probabilities and utilising that $P(x,y,z) = P(x|y,z)P(y,z) = P(y|x, z)P(x, z)$:
\begin{eqnarray}
P({\rm data} | S , \theta) P(S, \theta) &=& P(S | {\rm data}, \theta) P({\rm data}, \theta)\\
P({\rm data} | S , \theta) &=& \frac{ P(S | {\rm data}, \theta) P({\rm data}, \theta) }{ P(S, \theta) }\\
&=& \frac{ P(S | {\rm data}, \theta) P({\rm data} | \theta) P(\theta)}{ P(S | \theta)  P(\theta)}\\
&=& \frac{ P(S | {\rm data}, \theta) P({\rm data} | \theta)}{ P(S | \theta) } \label{eq:general1}
\end{eqnarray}
which is equal to the likelihood $\mathcal{L}$. \added{ At this point, our derivation depends on whether our selection effects are best modelled as a function of observables or as a function of latent parameters. In most cases, selection effects will be best modelled directly as a function of observables, and we would introducing an integral over all possible events $D$, so we can evaluate $P(S|\theta)$, 
\begin{eqnarray}
\mathcal{L}(\theta; {\rm data})&=& \frac{P(S|{\rm data},\theta) P({\rm data}|\theta)}{\int P(S, D|\theta)\, dD} \\
\mathcal{L}(\theta; {\rm data}) &=& \frac{P(S|{\rm data},\theta) P({\rm data}|\theta)}{\int P(S | D, \theta) P(D|\theta)\, dD}. \label{eq:app_main}
\end{eqnarray}
The second option, wherein selection effects can be more computationally efficiently modelled with the inclusion of latent parameters (for example, in the case where we do not have direct access to the observables upon which our data selection is determined), we can introduce our latent parameters in addition to an integral over all possible data:
\begin{eqnarray}
\mathcal{L}(\theta; {\rm data})&=& \frac{\int P(S|{\rm data}, L,\theta) P({\rm data}|L) P(L|\theta)\, dL}{\iint P(S|D,L,\theta)P(D|L,\theta) P(L|\theta)\, dD\, dL},\label{eq:latent}
\end{eqnarray}
where $L$ represents our latent parameters that model the true underlying values of our observables, such that our data is conditioned directly on them. In this formulation, the selection effects can depend both on data and latent variables.
}

\subsection{Supernova Selection Effects}
\label{app:selection2}
\added{
To turn the generalised equations from the previous sections into selection effects relevant for this model, need to highlight that $\theta$ represents only our top level parameters ($\Omega_m$, $w$, $\alpha$, $\beta$, etc), and that our parametrisation of true underlying values (for example $m_B$) takes the form of latent parameters. We thus continue from equation \eqref{eq:latent} and write out denominator $d$:
\begin{equation}
d = \iint  P(S|D, L, \theta) P(D | L, \theta) P(L | \theta)\, dD\, dL.
\end{equation}
In our formulation, we assume that our selection effects can be sufficiently encapsulated by latent parameters. That is, we simplify $P(S|D,L, \theta) \rightarrow P(S|L)$. This allows us to isolate our integral $\int P(D|L,\theta)$ in the above equation, integrate it to unity, and come to 
\begin{equation}
d = \int  P(S|L) P(L | \theta)\, dL.
\end{equation}	
Here we can now move from the generic label $L$ to something specific to our model. Specifically, we assume that our selection effects can be quantified using the true apparent magnitude $m_B$ and redshift $z$, and so $L \rightarrow \lbrace m_B, z\rbrace$, or formally $P(S|D, L, \theta) = P(S|z) P(S|m_B)$. We do not write out other latent parameters found in the model as they do not impact selection effects --- we would simply find them integrated to unity. Note that this assumption does not capture biases engendered by
Poisson noise fluctuations.  
We advocate that future analyses with
higher statistical precision do use a precisely determined $P(S|D)$.
Writing out the latent variables, our denominator becomes
\begin{equation}
d = \int P(S|z) P(S|m_B)P(z, m_B | \theta)\, dz \, dm_B.  \label{eq:sel} 
\end{equation}
We can express $P(z, m_B | \theta)$ as  $P(m_B | z, \theta) P(z | \theta)$, where the first term requires us to calculate the magnitude distribution of our underlying population at a given redshift, and the second term is dependent on survey geometry and supernovae rates. We can thus state}
\begin{equation}
d = \int \left[ \int P(S|m_B) P(m_B | z, \theta)\, dm_B \right] P(S|z)P(z|\theta)\, dz.
\end{equation}
By assuming that the distribution $P(S|z)P(z|\theta)$ is well sampled by the observed supernova redshifts, we can approximate the integral over redshift by evaluating
\begin{equation}
\int P(S|m_B) P(m_B | z, \theta)\, dm_B \label{eq:selint}
\end{equation}
for each supernova in the dataset -- i.e. Monte Carlo integration with assumed perfect importance sampling.

As stated in Section \ref{sec:selection}, the underlying population in apparent magnitude, when we discard skewness, can be represented as $\mathcal{N}(m_B|m_B^*(z), \sigma^*_{m_B})$, where
\begin{eqnarray}
m_B^*(z) &=& \langle M_B \rangle + \mu(z) - \alpha \langle x_1(z) \rangle + \beta \left(\langle c(z) \rangle + \sqrt{\frac{2}{\pi}}\sigma_c \delta_c\right)\label{eq:disc1} \\
\sigma^*_{m_B} &=& \sigma_{M_B}^2 + (\alpha \sigma_{x_1})^2 +  \left(\beta \sigma_c \sqrt{1 - \frac{2\delta_c^2}{\pi}}\right)^2. \label{eq:disc2}
\end{eqnarray}
Then, modelling $P(S|m_B)$ as either a normal or a skew normal, we can analytically perform the integral in equation \eqref{eq:selint} and reach equations \eqref{eq:seldes} and \eqref{eq:sellow-z}.

\subsection{Approximate Selection Effects}
\label{app:approx}

In this section, we investigate the effect of approximating the skew normal underlying color distribution as a normal. Specifically, equations \eqref{eq:disc1} and \eqref{eq:disc2} make the assumption that, for our color distribution, $\mathcal{N}^{\rm Skew}(\mu, \sigma, \alpha)$ is well approximated by $\mathcal{N}(\mu, \sigma)$. We sought to improve on this approximation by adjusting the mean and standard deviation of the approximated normal to more accurately describe the actual mean and standard deviation of a skew normal. With $\delta \equiv \alpha/\sqrt{1+\alpha^2}$, the correct mean and standard deviation are
\begin{eqnarray}
\mu_1 &=& \mu_0 + \sqrt{\frac{2}{\pi}} \delta \sigma_0 \\
\sigma_1 &=& \sigma_0 \sqrt{1 - \frac{2 \delta^2}{\pi}},
\end{eqnarray}
where we highlight that $\mu$ here represents the mean of the distribution, not distance modulus. We can then test the approximation $\mathcal{N}^{\rm Skew}(\mu_0, \sigma_0, \alpha) \rightarrow \mathcal{N}(\mu_1, \sigma_1)$. Unfortunately, this shift to the mean and standard deviation of the normal approximation where we treat $m_B$, $x_1$, and $c$ as a multivariate skew normal did not produce stable posterior surfaces. Due to this, we treat the underlying $m_B$, $x_1$, and $c$ populations as independent. 

\begin{figure*}
	\begin{center}
		\includegraphics[width=\textwidth]{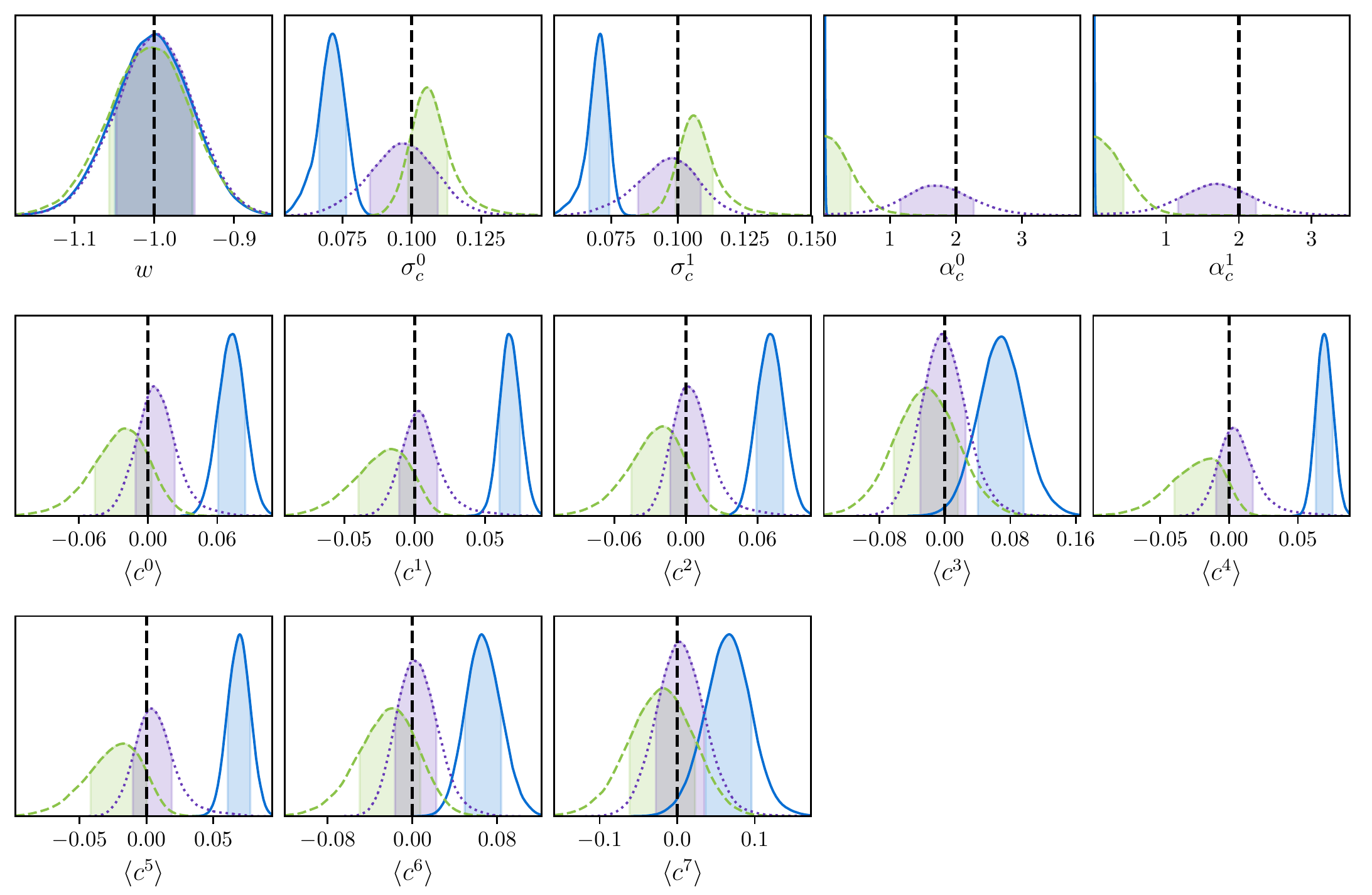}
	\end{center}
	\caption{Marginalised probability distributions for 100 realisations of cosmology, fit to Flat $w$CDM with prior $\Omega_m \sim \mathcal{N}(0.3, 0.01)$, each containing 1000 simulated high-$z$ and 1000 simulated low-$z$ supernovae. The dashed green surfaces represent a fit to an underlying Gaussian color population with the unshifted model. The blue solid surface represents fits to a skewed color population with the unshifted model, and the purple dotted surface represents a fit to a skewed color population with the shifted model. The superscript $0$ and $1$ denote the two different surveys (high-$z$ and low-$z$ respectively), and similarly the first four $\langle c^i \rangle$ parameters represent the four redshift nodes in the high-$z$ survey, and the last four represent the nodes for the low-$z$ survey. We can see that the shifted model is far better able to recover skewed input populations than the unshifted, performing better in terms of recovering skewness $\alpha_c$, mean color $\langle c \rangle$ and width of the color distribution $\sigma_c$. The unshifted model recovers the correct color mean and width if you approximate a skew normal as a normal: $\Delta\mu = \sqrt{2/\pi}\sigma_c\delta_c \approx 0.071$, which is approximately the deviation found in fits to the color population mean. Importantly, the unshifted model when run on skewed data (the solid blue) shows extreme bias in $\alpha_c$, where it fits strongly around zero regardless, showing it to be a poor approximation. Based on these results and the good performance in correctly recovering underlying populations of the shifted normal approximation, we adopt the shifted normal approximation in our model.}
	\label{fig:simple_w_super}
\end{figure*}

We tested a fixed $\sigma_c$ in the shift correction, such that $\mu_1 = \mu_0 + \sqrt{2/\pi}\delta k$, where we set $k=0.1$ to mirror the width of the input simulation population. This resulted in stable posterior surfaces, however this introduced recovery bias in several population parameters, and so we do not fix $\sigma_c$. Comparing whether we shift our normal in the approximation or simply discard skewness, Figure \ref{fig:shift} shows that the calculated efficiency is significantly discrepant to the actual efficiency if the normal approximation is not shifted. The biases when using shifted or unshifted normal approximations when we fit our model on Gaussian and skewed underlying populations are shown in Figure \ref{fig:simple_w_super}, and only the shifted normal approximation correctly recovers underlying population parameters.

\section{Numerical Optimisations}
\label{app:optimisations}

Not many fitting methodologies and algorithms can handle the thousands of fit parameters our model requires. By using Stan, we are able to take advantage automatic differentiation and the NUTS sampler, which is a class of Hamiltonian Monte Carlo samplers. Even with these advantages, early implementations of our model still had excessive fit times, with our desired sub-hour running time far exceeded. 

The simplest and most commonly found optimisation we employed was to precompute as much as possible. This is in a bid to reduce the complexity of the mathematical graph our model is translated into by Stan, to simplify the computation of surface derivatives. For example, when computing the distance modulus, redshift is encountered to various powers. Instead of computing those powers in Stan, we simply pass in several arrays of redshift values already raised to the correct power. Small changes like this however only give small improvements.

The primary numerical improvement we made on existing frameworks was to remove costly probability evaluations of multivariate normals. To increase efficiency, the optimum way to sample a multivariate normal is to reparameterise it such that instead of sampling $\mathcal{N}(\vec{x}|\vec{\mu}, \Sigma)$, we sample $\mathcal{N}(\vec{\delta}|0,1)$ where $\vec{x} = \vec{\mu} + L \vec{\delta}$ and $L$ is the cholesky decomposition of $\Sigma$. In this way, we can efficiently sample the unit normal probability distribution instead of sampling a multivariate normal probability distribution. Switching to this parametrisation resulted in a computational increase of an order of magnitude, taking fits for a sample of approximately 500 supernovae from roughly four hours down to thirty minutes. 

This parametrisation does come with one significant downside --- inflexibility. For each step the algorithm takes, we do not recompute the cholesky decomposition of the covariance of the summary statistics --- that happens once at the beginning of the model setup. If we had kept the full covariance matrix parametrisation we could modify the matrix easily --- for example when incorporating intrinsic dispersion we could simply add on a secondary matrix to create an updated covariance. However as the cholesky decomposition of a sum of matrices is not equal to the sum of the cholesky decomposition of each individual matrix, we would need to recompute the decomposition for each step, which discards most of the computational benefit just gained.

Considering a $3\times3$ matrix with cholesky decomposition
\begin{equation}
L = \begin{pmatrix}
a & 0 & 0 \\ b & c & 0 \\ d & e & f \\
\end{pmatrix},
\end{equation}
the original covariance matrix $\Sigma$ is given by
\begin{equation}
\Sigma = \begin{pmatrix}
a^2 & ab & ad \\ ab & b^2 + c^2 & bd + ce \\ ad & bd + ce & d^2 + e^2 + f^2\\
\end{pmatrix}.
\end{equation}
Now, the primary source of extra uncertainty in the intrinsic dispersion models comes from chromatic smearing, which primarily influences the recovered color parameter, which is placed as the last element in the observables vector $\lbrace m_B, x_1, c\rbrace$. We can now see that it is possible to add extra uncertainty to the color observation on the diagonal without having to recompute the cholesky decomposition - notice that $f$ is unique in that it is the only element of $L$ that appears in only one position in the covariance matrix. To take our covariance and add on the diagonal uncertainty for color an extra $\sigma_e$ term, we get
\begin{equation}
C = \begin{pmatrix}
\sigma_{m_B}^2 & \rho_{0,1} \sigma_{m_B} \sigma_{x_1} & \rho_{0,2} \sigma_{m_B} \sigma_c \\
\rho_{0,1} \sigma_{m_B} \sigma_{x_1} & \sigma_{x_1}^2 & \rho_{1, 2} \sigma_{x_1} \sigma_c \\
\rho_{0,2} \sigma_{m_B} \sigma_c & \rho_{1, 2} \sigma_{x_1} \sigma_c &  \sigma_c^2 + \sigma_e^2 \\
\end{pmatrix}.
\end{equation}
The cholesky decomposition of this is, in terms of the original cholesky decomposition, is
\begin{equation}
L = \begin{pmatrix}
a & 0 & 0 \\ b & c & 0 \\ d & e & f + g \\
\end{pmatrix},
\end{equation}
where $g = \sqrt{f^2 + \sigma_e^2} - f$. This allows an easy update to the cholesky decomposition to add extra uncertainty to the independent color uncertainty. For both the {\gten} and {\celeven} models, we ran fits without the cholesky parametrisation to allow for extra correlated dispersion (instead of just dispersion on $c$), but find no decrease in bias or improved fit statistics, allowing us to use the more efficient cholesky parametrisation.

\end{document}